\documentclass[aps,prd,notitlepage,tightenlines,nofootinbib,twocolumn,superscriptaddress]{revtex4-2}

\usepackage{multirow}

\usepackage{amsmath}
\usepackage{amssymb,amsthm}
\usepackage{mathtools}
\usepackage{mathrsfs}
\usepackage{relsize}
\usepackage{bm}
\allowdisplaybreaks[4]  
\usepackage{epsfig,graphics,graphicx}
\usepackage{color,xcolor}
\usepackage{subfigure}
\usepackage{array}
\usepackage{ragged2e}
\usepackage{lineno}
\usepackage{natbib}
\usepackage{hyperref}
\hypersetup{colorlinks=true,
            linkcolor=blue,
            anchorcolor=blue,
            citecolor=magenta,
            filecolor=blue,
            urlcolor=blue,
            bookmarksnumbered=true,
            pdfview=FitB
}
\usepackage{enumitem}
\usepackage{ulem}
\usepackage{listings}
\usepackage{soul}
\usepackage{cancel}

\colorlet{darkgreen}{green!50!black}
\colorlet{brightyellow}{yellow!75!red}
\colorlet{orange}{red!50!yellow}
\colorlet{darkgray}{gray!50!black}
\colorlet{darkred}{red!50!black}
\def\dd{{\mathrm{d}}}

\newcommand{\half}[1][1] {\mathsmaller{\frac{#1}{2}}}

\makeatletter
\newcommand*{\transpose}{%
  {\mathpalette\@transpose{}}%
}
\newcommand*{\@transpose}[2]{%
  \raisebox{\depth}{$\m@th#1\intercal$}%
}
\makeatother

\begin{document}

\title{Stress out of charmonia}

\author{Siqi Xu}
\affiliation{Institute of Modern Physics, Chinese Academy of Sciences, Lanzhou 730000, China}
\affiliation{School of Nuclear Science and Technology, University of Chinese Academy of Sciences, Beijing 100049, China}

\author{Xianghui Cao}
\affiliation{Department of Modern Physics, University of Science and Technology of China, Hefei, Anhui 230026, China}

\author{Tianyang Hu}
\affiliation{Department of Modern Physics, University of Science and Technology of China, Hefei, Anhui 230026, China}

\author{Yang Li}
\affiliation{Department of Modern Physics, University of Science and Technology of China, Hefei, Anhui 230026, China}
\affiliation{Department of Physics and Astronomy, Iowa State University, Ames, Iowa 50010, U.S.}

\author{Xingbo Zhao}
\affiliation{Institute of Modern Physics, Chinese Academy of Sciences, Lanzhou 730000, China}
\affiliation{School of Nuclear Science and Technology, University of Chinese Academy of Sciences, Beijing 100049, China}
\affiliation{CAS Key Laboratory of High Precision Nuclear Spectroscopy, Institute of Modern Physics, Chinese Academy of Sciences, Lanzhou 730000, China}

\author{James P. Vary}
\affiliation{Department of Physics and Astronomy, Iowa State University, Ames, Iowa 50010, U.S.}

\date{\today}

\begin{abstract}

We investigate the gravitational form factors of charmonium. Our method is based on a Hamiltonian formalism on the light front known as basis light-front quantization. The charmonium mass spectrum and light-front wave functions were obtained from diagonalizing an effective Hamiltonian that incorporates confinement from holographic QCD and one-gluon exchange interaction from light-front QCD. We proposed a quantum many-body approach to construct the hadronic matrix elements of the energy momentum tensor $T^{++}$ and $T^{+-}$, which are used to extract the gravitational form factors $A(Q^2)$ and $D(Q^2)$. The obtained form factors satisfy the known constraints, e.g. von Laue condition. From these quantities, we also extract the energy, pressure and light-front energy distributions of the system. We find that hadrons are multi-layer systems. 

\end{abstract}
\maketitle

\section{Introduction}

The year 2023 marked the 50 year anniversary of quantum chromodynamics (QCD). However, our understanding of the strong interaction remains incomplete \cite{Gross:2022hyw}. 
One of the challenges is the strong-coupling nature of QCD. As such, the strong force within hadrons is dramatically different from what the QCD Lagrangian naïvely tells us, in contrast to weakly coupled systems, e.g. the hydrogen atom. 
The primary operators that describe the distribution of stress within a system are the energy-momentum tensor (EMT) $T^{\mu\nu}$ \cite{Pagels:1966zza}.  These operators also dictate how the system gravitates, and conversely, how it responds to the gravitational field \cite{Kobzarev:1962wt}. The access to the hadronic matrix elements (HME) of the EMT has been a hot topic in recent literature. %
Kumano et al. extracted the EMT of the pion from the process $\gamma^*\gamma \to \pi^0\pi^0$ measured on Belle \cite{Kumano:2017lhr}. Burkert et al. obtained the stress inside the proton using deeply virtual Compton scattering data and deeply vector meson production data collected from Jefferson Lab \cite{Burkert:2018bqq, Burkert:2021ith, Kumericki:2019ddg, Dutrieux:2021nlz, Burkert:2023atx}. Duran et. al. accessed the gluon contributions for the proton using near-threshold  electroproduction of $J/\psi$ \cite{Duran:2022xag}. The precision of these experiments is expected to dramatically improve in the forthcoming era of electron-ion colliders \cite{Accardi:2012qut, Anderle:2021wcy, Accardi:2023chb}. 
Lattice simulation of the nucleon and the pion EMT are recently available with a pion mass $M_\pi = 170 \, \mathrm{MeV}$ \cite{Hackett:2023rif, Hackett:2023nkr} (cf.~\cite{Yang:2018nqn, Shanahan:2018pib, Shanahan:2018nnv, Pefkou:2021fni}). The hadronic EMT were also investigated in various theories and phenomenological models, e.g., \cite{Polyakov:2002yz, Cebulla:2007ei, Mai:2012cx, Mai:2012yc, Jung:2013bya, Cantara:2015sna, Hudson:2017oul, Polyakov:2018exb, Anikin:2019kwi, Ozdem:2019pkg, Varma:2020crx, Chakrabarti:2020kdc, Metz:2021lqv, More:2021stk, Tong:2022zax, Mamo:2022eui, Fujita:2022jus, Xing:2022mvk, Ozdem:2022zig, Freese:2022jlu, More:2023pcy, Pasquini:2023aaf, GarciaMartin-Caro:2023klo, Chakrabarti:2023djs, Amor-Quiroz:2023rke, Guo:2023pqw, Kou:2023azd, Dehghan:2023ytx, Cao:2023ohj, Won:2023zmf, Krutov:2023ztx, Li:2023izn, Amor-Quiroz:2023rke, Alharazin:2023uhr, Hagiwara:2024wqz, Nair:2024fit}. See Refs.~\cite{Polyakov:2018zvc, Burkert:2023wzr} for recent reviews.   

One of the challenges to access the hadronic EMT is that $T^{\mu\nu}$ contains the interaction, which is strong at low energy. In this work, we present the first calculation of the charmonium EMT. Charmonium, first discovered 50 years ago in the \textit{November revolution}, is an ideal system to probe the properties of the strong force. This system consists of a pair of charm and anti-charm quarks. Since $m_c \gg \Lambda_\text{QCD}$ and the strong coupling $\alpha_s(m_c) \ll 1$, phenomena occurring at the scale $\sim m_c$ may be treated perturbatively. On the other hand, with the binding energy $\alpha_s m_c \gtrsim \Lambda_\text{QCD}$,  a non-perturbative treatment is required to describe the properties of the QCD bound states including the stress within. Hence, this bound system involves the interplay of the perturbative and non-perturbative physics of QCD. It is sometimes dubbed as the ``hydrogen atom of QCD" \cite{Brambilla:2010cs}. 
Our method is based on basis light-front quantization (BLFQ) \cite{Vary:2009gt}. This method stems from the light-front Hamiltonian formalism, which finds its roots from two lineages \cite{Namyslowski:1985zq, Zhang:1994ti, Burkardt:1995ct, Brodsky:1997de, Carbonell:1998rj, Heinzl:2000ht, Miller:2000kv, Bakker:2013cea, Hiller:2016itl, Brodsky:2022fqy}. One is Dirac \cite{Dirac:1949cp} and Weinberg's \cite{Weinberg:1966jm} efforts to simplify relativistic dynamics, which leads to the discovery of the front form of quantum field theory \cite{Chang:1968bh, Bardakci:1968zqb, Kogut:1969xa, Tiktopoulos:1969wp, Chang:1972xt, Yan:1973qg, Tomboulis:1973jn, Rohrlich:1973ij, Casher:1976ae, Nakanishi:1976vf, Leutwyler:1977vy, Thorn:1978kf, Thorn:1979gv, Franke:1981ma}. The other is the light-cone current algebra \cite{Gell-Mann:1960mvl, Gell-Mann:1962yej, Gell-Mann:1964ish, Fubini:1964boa, Wilson:1969zs, Fritzsch:1971xe} and Feynman's parton model \cite{Feynman:1969ej, Bjorken:1969ja, LlewellynSmith:1970ik, Feynman:1971wr, Feynman:1973xc} emerging in deep inelastic scattering \cite{Kogut:1972di, Lepage:1980fj}. These two complementary perspectives are unified in the light-front Hamiltonian formalism, which allows us to tackle QCD as a quantum many-body problem, as described by discretized light-cone quantization (DLCQ) \cite{Pauli:1985ps} and BLFQ \cite{Vary:2009gt}. 

BLFQ was successfully applied to a number of systems including charmonium,  and to a number of hadronic observables including mass spectra, radiative transitions, parton distributions and transverse momentum distributions \cite{Honkanen:2010rc, Zhao:2014xaa, Wiecki:2014ola, Hu:2020arv, Nair:2022evk, Nair:2023lir, Li:2021jqb, Li:2022izo, Li:2022ytx, Li:2023izn, Li:2015zda, Li:2017mlw, Leitao:2017esb, Li:2018uif, Adhikari:2018umb, Tang:2018myz, Lan:2019img, Tang:2019gvn, Tang:2020org, Li:2021ejv, Li:2021cwv, Wang:2023nhb, Jia:2018ary, Lan:2019vui, Lan:2019rba, Qian:2020utg, Mondal:2021czk, Adhikari:2021jrh, Li:2022mlg, Zhu:2023lst, Mondal:2019jdg, Xu:2021wwj, Liu:2022fvl, Hu:2022ctr, Peng:2022lte, Zhu:2023lst, Zhang:2023xfe, Zhu:2023nhl, Kuang:2022vdy, Lan:2021wok, Xu:2022abw, Xu:2023nqv, Lin:2023ezw, Kaur:2024iwn}.
For charmonium, the effective Hamiltonian combines confinement from holographic light-front QCD \cite{Brodsky:2014yha} for the long-distance physics as well as one-gluon exchange interaction from light-front QCD \cite{Wiecki:2014ola} for the short-distance physics \cite{Li:2015zda}. The computed mass spectrum is shown to be within 40 MeV deviation with the PDG compilation \cite{Li:2017mlw, Tang:2018myz, Tang:2019gvn, Tang:2020org}. For the ground state $\eta_c$, the mass deviation is about $\sim 15\%$ of the total binding energy. The obtained light-front wave functions (LFWFs) were used to investigate a variety of observables, including the decay constants \cite{Li:2015zda, Li:2017mlw}, electromagnetic form factors \cite{Adhikari:2018umb}, parton distribution functions \cite{Lan:2019img}, and GPDs \cite{Adhikari:2018umb}. In particular, the parameter-free predictions of the radiative transition widths and radiative transition form factors are in remarkable agreement with the experimental measurements whenever available \cite{Li:2021ejv, Li:2021cwv, Wang:2023nhb}. 

The rest of the article is organized as follows. Sect.~\ref{sect:light-front Hamiltonian formalism} briefly introduces the light-front Hamiltonian formalism including the basis function representation chosen for BLFQ. Sect.~\ref{sect:EMT} describes the energy-momentum tensor and the light-front densities associated with these operators. The light-front wave function representation of the gravitational form factors $A(Q^2)$ and $D(Q^2)$ are presented in Sect.~\ref{sect:LFWF_representation}. We present the numerical results in Sect.~\ref{sect:results}. Finally, we conclude in Sect.~\ref{sect:summary}. 

\section{Hamiltonian field theory on the light front}\label{sect:light-front Hamiltonian formalism}

In light-front QCD, the quantum state of a hadron $|\psi_h\rangle$ can be obtained by solving the light-front Schrödinger equation \cite{Brodsky:1997de}, 
\begin{equation}
i\frac{\partial}{\partial x^+}|\psi_h(x^+)\rangle = \frac{1}{2}P^-|\psi_h(x^+)\rangle\,.
\end{equation}
Here, we adopt the light-front coordinates, $x^\pm = x^0 \pm x^3$, $\vec x_\perp = (x^1, x^2)$.
The hadronic state vector $|\psi_h(p, j, m_j)\rangle$ can be further classified by 
its momentum $p^\mu$, total spin $j$ and spin magnetic projection $m_j$, viz. \cite{Weinberg:1995mt, Brodsky:1997de},
\begin{align}
& P^\mu |\psi_h(p, j, m_j)\rangle = p^\mu |\psi_h(p, j, m_j)\rangle, \label{eqn:ev_P} \\
& \vec S^2  |\psi_h(p, j, m_j)\rangle = j(j+1) |\psi_h(p, j, m_j)\rangle, \label{eqn:ev_S2}\\
& S^+ |\psi_h(p, j, m_j)\rangle = m_j |\psi_h(p, j, m_j)\rangle, \label{eqn:ev_S+}
\end{align}
where, $P^\mu$ is the 4-momentum operator, $\vec S^2 = W^2/P^2$ is the total spin operator, constructed from the Pauli-Lubanski vector $W^\mu = \frac{1}{2}\varepsilon^{\mu\nu\rho\sigma}P_\nu J_{\rho\sigma}$,  where $J^{\alpha\beta}$ is the generalized angular momentum operator, i.e., the generator of the Lorentz symmetry, and $S^+ = W^+/P^+$ is the magnetic projection of the total spin operator, also known as the light-front helicity operator. The light-front energy for a free hadron is $p^- = (p^2_\perp+M^2)/p^+$, where $M$ is the invariant mass of the particle.

These operators are not all kinematical, i.e. some of these operators contain interactions. The dynamical nature of operators depends on the choice of the initial surface, i.e., the quantization scheme. Light-front quantization with initial surface $x^+ = 0$ turns out to be equipped with the maximal number (7) of kinematical operators out of the 10 Poincaré generators \cite{Dirac:1949cp}. The remaining 3 generators originate from the same local interaction in quantum field theory \cite{Carbonell:1998rj}, 
\begin{align}
P^\mu_\text{int} =\,& \omega^\mu \int \dd^4 x\, \delta(\omega\cdot x) \mathcal H_\text{int}(x), \label{eqn:P_int} \\
J^{\mu\nu}_\text{int} =\,& \int \dd^4 x\, \delta(\omega\cdot x)  (x^\mu \omega^\nu - \omega^\mu x^\nu) \mathcal H_\text{int}(x) \label{eqn:J_int} \,.
\end{align}
Here, $\omega^\mu = (\omega^+, \omega^-, \vec\omega_\perp) = (0, 2, \vec 0_\perp)$ is the null vector that indicates the orientation of the light-front quantization surface. 
On the other hand, the Poincaré generators stem from the Noether currents, e.g., 
\begin{equation}
P^\mu = \int \dd^3x\, T^{+\mu}(x) \,,
\end{equation}
where, $T^{\mu\nu}$ is the EMT. In particular, the density of the light-front interaction energy $\mathcal H_\text{int}$ is related to the $+-$ component of the EMT, $\mathcal H_\text{int} = \frac{1}{2}T^{+-}_\text{int}$.  
Momentum conservation (\ref{eqn:ev_P}) imposes constraints on the forward HMEs of the EMT. It is straightforward to show, 
\begin{equation}\label{eqn:forward_HME}
\langle \psi_h(p, j, m_j')| T^{+\mu}(0) | \psi_h(p, j, m_j)\rangle = 2p^+ p^\mu \delta_{m_j, m'_j}\,.
\end{equation}
Similar constraints can be obtained from the conservation of the angular momentum (\ref{eqn:ev_S2}--\ref{eqn:ev_S+}) as shown by Lorcé et. al. \cite{Lorce:2019sbq}. 

In QCD, the force is strong and we need a non-perturbative approach to solve the eigenvalue problems (\ref{eqn:ev_P}--\ref{eqn:ev_S+}). 
Inspired by the success of \textit{ab initio} nuclear structure theory, especially the no-core shell model (NCSM) \cite{Barrett:2013nh}, 
Vary et. al. proposed BLFQ as a non-perturbative computational framework to solve light-front QCD \cite{Vary:2009gt}. 
In BLFQ,  the light-front Schrödinger equation is cast into a matrix eigenvalue equation, 
\begin{equation}
\sum_j H_{ij} \psi_j = M^2  \psi_i\,,
\end{equation}
where $H_{ij} = \langle i | H | j \rangle$ and $\psi_i = \langle i | \psi\rangle$. Here $|i\rangle$ is the basis state and $|\psi\rangle$ is the state vector of a hadron, and $H = P^+P^- - \vec P_\perp^2$ is equivalent to the light-front Hamiltonian operator $P^-$ since $P^+$ and $\vec P^2_\perp$ are kinematical. The basis states are chosen to retain the maximal kinematical symmetries of the system. A convenient choice is the 2D harmonic oscillator function in the transverse direction and Jacobi polynomials in the longitudinal direction. The many-body basis is then constructed from the tensor product of the single-particle basis. Other choices are also available, such as the discretized momentum basis adopted in DLCQ \cite{Hornbostel:1988fb}, conformal basis functions employed in Hamiltonian conformal truncation method \cite{Anand:2020gnn}, and coherent basis in light-front coupled cluster method \cite{Chabysheva:2011ed}. These bases have been shown successful in applications to low-dimensional or simpler quantum field theories \cite{Hiller:2016itl}. 
 
In BLFQ, one adopts an effective Hamiltonian suitable for the finite basis space with low resolution.  Two approaches are adopted. In the top-down approach, one starts from QCD and perform a similarity renormalization group evolution \cite{Glazek:1993rc, Wilson:1994fk, Brisudova:1996vw, Brisudova:1997rv, Glazek:2017rwe}. In the bottom-up approach, one starts with a phenomenological effective Hamiltonian that embodies the key physics of the system \cite{Li:2015zda, Li:2017mlw}. The latter is similar to the nuclear shell model in nuclear structure calculations, where the role of the first approximation, the nuclear shell model, is played by holographic light-front QCD, as shown by Brodsky and de Téramond \cite{Brodsky:2014yha}. 

In this work, we adopt the phenomenological effective Hamiltonian with a confining interaction from light-front holographic QCD. The form is remarkably simple, $V_{q\bar q} \propto x(1-x) r^2_\perp$, where, $x=p^+_q/P^+_h$ is the longitudinal momentum fraction of the quark, and $\vec r_\perp$ is the relative transverse separation of the quark and antiquark. This potential generates a 2D harmonic oscillator wave functions as the eigenfunction, which coincides with the basis function in BLFQ in the transverse direction which provides computational advantages such as  enabling factorization of the center-of-mass motion and facilitation of the transformation between relative and single-particle coordinates. 
A longitudinal interaction of the 't Hooft type is supplemented to generate confinement in the longitudinal direction \cite{Chabysheva:2012fe, Li:2021jqb, deTeramond:2021yyi, Li:2022izo}. 
We further supplement confinement with a one-gluon exchange interaction obtained from light-front QCD. This piece of interaction is essential for describing the short-distance physics as well as the spin structure \cite{Wiecki:2014ola}. 

The basis functions are the eigen-functions of the confining interactions. Their analytic forms are, 
\begin{align}
\phi_{nm}(\vec v_\perp) =\,&N_{nm} \kappa^{-1}  \Big(\frac{v_\perp}{\kappa}\Big)^{|m|} e^{i m \theta - \frac{v^2_\perp}{2\kappa^2}} L_n^{|m|}\Big(\frac{v^2_\perp}{\kappa^2}\Big),  \label{eqn:HO}\\
\chi_l(x) =\,& N_l x^{\frac{\beta}{2}}(1-x)^{\frac{\alpha}{2}} P_l^{(\alpha,\beta)}(2x-1), \label{eqn:Jacobi}
\end{align}
 where, $N_l = \sqrt{4\pi(2l+\alpha+\beta+1)} \sqrt{\frac{\Gamma(l+1)\Gamma(l+\alpha+\beta+1)}{\Gamma(l+\alpha+1)\Gamma(l+\beta+1)}}$, $N_{nm} = \sqrt{\frac{4\pi n!}{(n+|m|)!}}$, $\vec v_\perp = \vec p_\perp/\sqrt{x(1-x)}$ is the holographic momentum, $v_\perp = |\vec v_\perp|$, and $\theta = \arg \vec v_\perp$, $x = p^+/P^+$ is the longitudinal momentum fraction. The basis parameters $\alpha = {2m_{\bar q}(m_q + m_{\bar q})}/{\kappa^2}$, $\beta = {2m_{q}(m_q + m_{\bar q})}/{\kappa^2}$, where $\kappa$ is the strength of the confining interaction and $m_q = m_{\bar q} = m_c$ is the effective quark mass.   
Then effective Hamiltonian for the $c \bar{c}$ system is, 
\begin{multline}
H_\text{eff} = \sum_I h_{I} a^\dagger_I a_I +  \frac{1}{2}\sum_{I, J, I', J'} V_{IJI'J'} a^\dagger_I a^\dagger_J a_{J'} a_{I'}  \,.
\end{multline}
Here, $I, J, I', J'$ each represent a collection of single-particle quantum numbers: radial quantum number $n$, angular quantum number $m$, longitudinal quantum number $l$, spin $s$, color $i$ and parton flavor $f$, e.g., $I= \{n, m, l, s, i, f\}$. The one-body effective Hamiltonian $h_I = 4m_c^2 + 2\kappa^2(2n+|m|+l+\frac{3}{2}) + ({\kappa^4}/{4m_c^2}) l(l+1)$ incorporates the kinetic energy as well as the long-distance interactions, i.e. confinement. We adopted the one-gluon exchange interaction as the two-body effective interaction, which is the dominant physics at short distance for charmonium. %
For charmonium, we further impose a valence approximation, i.e. keeping only the $c\bar c$ Fock sector. 

BLFQ adopts the $N_\text{max}$ basis truncation scheme to render the basis space finite. Specifically for charmonium, we impose the conditions, 
\begin{equation}
 2n+|m|+1 \le N_{\text{max}}, \quad 0 \le l \le L_{\text{max}}.
\end{equation} 
$N_\text{max}$ defines the UV and IR regulators of the basis space as $\Lambda_\textsc{uv} = \sqrt{N_\text{max}}\kappa$, and $\lambda_\textsc{ir} = \kappa/\sqrt{N_\text{max} }$. $L_\text{max}$ defines the resolution of the longitudinal momentum fraction $\delta x = L^{-1}_\text{max}$.
For simplicity, we have tied the basis scale parameter to the confining strength $\kappa$. 

We diagonalized the Hamiltonian with different $N_\text{max}, L_\text{max}$ to obtain the mass squared eigenvalues and the LFWFs. The model parameters $\kappa$, and $m_c$ were fit to the charmonium mass spectrum below the $D\overline D$ threshold, which is a good approximation for states with narrow resonance. 
Since radiative corrections are neglected, we adopt $N_\text{max} = L_\text{max} = 8$ for observables that are sensitive to the short-distance quantum fluctuations, which corresponds to a UV resolution $\Lambda_\textsc{uv} \approx M_{c\bar c}$. To estimate the uncertainty associated with the basis resolution, we quote the difference between the $N_\text{max} = L_\text{max} = 8$ results and the $N_\text{max} = L_\text{max} = 16$ results. 
The details of the model can be found in Ref.~\cite{Li:2017mlw}. 

From the basis functions, we can reconstruct the valence LFWFs,
\begin{multline}\label{eqn:basis_reprensetation}
\psi^{(m_j)}_{s\bar s/h}(x, \vec k_\perp) = 
\sum_{n,m,l} \psi^{(m_j)}_{s\bar s/h}(n, m, l) \\
\times \phi_{nm}(\vec k_\perp/\sqrt{x(1-x)}) \chi_{l}(x)\,,
\end{multline}
where, $\psi^{(m_j)}_{s\bar s/h}(n, m, l) $ are the basis coefficients obtained from the diagonalization. 
The state vector can be represented by the LFWFs as, 
\begin{multline}
|\psi_h(P, j, m_j)\rangle = \sum_{s, \bar s} \int_0^1 \frac{\dd x}{2x(1-x)} \int \frac{\dd^2k_\perp}{(2\pi)^3} \\
\times \psi_{s\bar s/h}^{(m_j)}(x, \vec k_\perp)  \frac{1}{\sqrt{N_c}} \sum_{i=1}^{N_c} b^\dagger_{si}\big(xP^+, \vec k_\perp+x\vec P_\perp\big) \\
\times d^\dagger_{\bar si}\big((1-x)P^+, -\vec k_\perp+(1-x)\vec P_\perp\big) |0\rangle.
\end{multline}
Here, $P^\mu = (P^-, P^+, \vec P_\perp)$ is the 4-momentum of the bound state $h$. 
$\vec k_\perp = \vec p_\perp - x\vec P_\perp$ is the relative transverse momentum, and $x = p^+ / P^+$ is the longitudinal momentum fraction. 
The charmonium LFWFs are available in Mendeley Data repository \cite{Li:2019}. 


\section{Energy-momentum tensor} \label{sect:EMT}

By virtue of Poincaré symmetry, the HMEs of the EMT for a spin-0 particle, such as $\eta_c$ and $\chi_{c0}$, can be parametrized in terms of covariant tensors \cite{Polyakov:2018zvc}, 
\begin{multline}
\langle P+\half q | T^{\mu\nu}(0) | P - \half q \rangle = \\
2P^\mu P^\mu A(-q^2) + \frac{1}{2}(q^{\mu}q^\nu - q^2 g^{\mu\nu}) D(-q^2)\,,
\end{multline} 
where Lorentz scalar functions $A(-q^2)$ and $D(-q^2)$ are known as the gravitational form factors (GFFs). 

The 2D Fourier transform of the HMEs in the Drell-Yan frame $q^+ = 0$ can be interpreted as the hadronic EMT on the transverse plane \cite{Lorce:2018egm, Freese:2021czn, Freese:2021qtb, Freese:2021mzg, Li:2022hyf, Freese:2022fat, Cao:2023ohj, Freese:2023abr},
\begin{equation}
\mathcal T^{\alpha\beta}(\vec r_\perp;  P) = \int \frac{\dd^2q_\perp}{(2\pi)^2} e^{-i\vec q_\perp \cdot \vec r_\perp} 
t^{\alpha\beta}(\vec q_\perp;  P) \,,
\end{equation}
where, $t^{\alpha\beta}(\vec q_\perp; P)$ is the normalized hadronic matrix elements, 
\begin{equation}
t^{\alpha\beta}(\vec q_\perp; P) =  \frac{1}{2P^+}\langle P+\frac{1}{2}q|T^{\alpha\beta}(0)|P - \frac{1}{2}q\rangle \,.
\end{equation}
The factor $1/(2P^+)$ is due to the normalization of the state vector:
\begin{equation}
\langle p' | p \rangle = 2p^+(2\pi)^3\delta(p^+-p'^+)\delta^2(p_\perp - p'_\perp)\,.
\end{equation}

Note that $t^{\alpha\beta}$ (hence $\mathcal T^{\alpha\beta}$) depends on the c.m. momentum $P^\mu$ of the hadron \cite{Freese:2021mzg}. 
The density of the light-front longitudinal momentum $P^+$ is \cite{Abidin:2008sb, Lorce:2018egm}, 
\begin{equation}
\mathcal T^{++}(\vec r_\perp; P) 
= P^+\int  \frac{\dd^2q_\perp}{(2\pi)^2} e^{-i\vec q_\perp \cdot \vec r_\perp} 
A(q_\perp^2)\,.
\end{equation} 
Similarly, the density of the light-front transverse momentum $\vec P_\perp$ is,
\begin{equation}
\mathcal T^{+i}(\vec r_\perp; P)  = P^i_\perp\int  \frac{\dd^2q_\perp}{(2\pi)^2} e^{-i\vec q_\perp \cdot \vec r_\perp} 
A(q_\perp^2)\,.
\end{equation} 
From these expressions, the Fourier transform of GFF $A(q_\perp^2)$, 
\begin{equation}
\mathcal A(r_\perp) = \int  \frac{\dd^2q_\perp}{(2\pi)^2} e^{-i\vec q_\perp \cdot \vec r_\perp} 
A(q_\perp^2),
\end{equation}
can be interpreted as the matter density. In Refs.~\cite{Abidin:2008sb, Lorce:2018egm, Freese:2022fat}, this density is called the internal light-front longitudinal momentum density. We emphasize that $\mathcal A(r_\perp)$ describes the convective distribution of all three momenta. 
Note that for spin-$\frac{1}{2}$ particles, such as the nucleon, there is an extra contribution from spin, 
\begin{equation}
\mathcal T^{+i}_{ss}(\vec r_\perp; P) = P^i_\perp \mathcal A(r_\perp) + \big(\vec\nabla \times \vec{\mathcal S}(r_\perp)\big)^i_\perp\,.
\end{equation} 
Here, $\vec{\mathcal S}(r_\perp) = 2  \mathcal J(r_\perp)\vec s$ is the spin density, $\vec s = s \hat z$, and $\mathcal J(r_\perp)$ is the 2D Fourier transform of the GFF $J(Q^2)$, which exists for spin-$\frac{1}{2}$ particles.  N.B. $\mathcal J(r_\perp)$ is normalized to 1/2. 
The spin current $\vec\nabla \times \vec{\mathcal S}$ is non-convective. 

Momentum conservation (\ref{eqn:forward_HME}) implies the conservation of matter,
\begin{equation}\label{eqn:A_forward}
A(0) = 1 \; \Leftrightarrow \; \int\dd^2r_\perp \, \mathcal A(r_\perp) = 1.
\end{equation}
The 3D r.m.s radius of the matter distribution is, 
\begin{equation}
r^2_A  \equiv \frac{3}{2} \int\dd^2r_\perp \, r^2_\perp \, \mathcal A(r_\perp) = -6A'(0).
\end{equation}

The distribution of the light-front energy $P^- = (P^2_\perp+M^2)/P^+$ in a hadron state with mass $M$ reads \cite{Cao:2023ohj}, 
\begin{equation}\label{eqn:LF_energy_distribution}
\mathcal T^{+-}(r_\perp; P) = \frac{P^2_\perp \mathcal A(r_\perp) + \mathcal M^2(r_\perp)}{P^+}\, 
\end{equation} 
where, 
\begin{align}
 \mathcal M^2(r_\perp) =\,& \int  \frac{\dd^2q_\perp}{(2\pi)^2} e^{-i\vec q_\perp \cdot \vec r_\perp} M^2(q^2_\perp), \label{eqn:M2_density} \\
M^2(q^2_\perp) =\,& \Big(M^2+\frac{1}{4}q^2_\perp\Big)A(q^2_\perp) + \frac{1}{2}q_\perp^2 D(q^2_\perp). \label{eqn:M2FF}
 \end{align}
The first term ${\mathcal A(r_\perp)P^2_\perp }/{P^+}$ represents the light-front energy due to the motion of the particle. 
Here, we have introduced a new form factor $M^2(q^2_\perp)$ and its 2D light-front density $\mathcal M^2(r_\perp)$. In Refs.~\cite{Lorce:2018egm, Freese:2022fat}, this density [Eq.~(\ref{eqn:M2_density})] is termed the internal light-front energy density. We find it is better to interpret $\mathcal M^2(r_\perp)$ as the distribution of the invariant mass squared. Hereafter, we will use these two terms interchangeably. 
For spin-$\frac{1}{2}$ particles, the spin term also contributes to the light-front energy density $\mathcal T^{+-}(r_\perp)$, 
\begin{multline}
\mathcal T^{+-}_{ss}(r_\perp; P) 
=  \\
\frac{P^2_\perp \mathcal A(r_\perp) + \mathcal M^2(r_\perp) + \vec P_\perp \cdot (\vec\nabla \times \vec{\mathcal S}(r_\perp)) }{P^+} \,.
\end{multline}
The last term vanishes upon integration. 

Light-front energy conservation (\ref{eqn:forward_HME}) implies \cite{Cao:2023ohj}, 
\begin{equation}\label{eqn:M2_forward}
M^2(0) = M^2 \; \Leftrightarrow \; \int\dd^2r_\perp\,  \mathcal M^2(r_\perp) = M^2.
\end{equation}
From (\ref{eqn:M2FF}), the above condition is equivalent to 
\begin{equation}\label{eqn:q2D_forward}
\lim_{q_\perp \to 0} q_\perp^2 D(q_\perp^2) = 0.
\end{equation}
As we will see later, this forward limit constraint, also known as the von Laue condition, is a necessary condition for the force equilibrium of a hadron \cite{Hudson:2017oul, Polyakov:2018zvc}. 
  It is useful to define the r.m.s.~radius of the invariant mass squared distribution as, 
 \begin{align}
 r^2_{M^2} =\,& \frac{3}{2M^2}\int \dd^2 r_\perp \, r_\perp^2 \, \mathcal M^2(r_\perp) \\
 =\,& r_A^2 - \frac{3}{2}\lambda^2_C (1+2D)\,.
  \end{align}
Here, $\lambda_C = 1/M$ is the Compton wavelength of the hadron. 
$D = D(0)$ is called the $D$-term, also called the Druck term, and is dubbed as the last global unknown of hadrons \cite{Polyakov:2018zvc}. 

The above definitions show the advantage of light-front formalism in extracting frame-independent hadronic densities. In general, the 
hadronic EMT can be parametrized by (the proper) energy density $\mathcal E(x)$, pressure $\mathcal P(x)$ and (traceless) shear $\Pi^{\alpha\beta}(x)$ as \cite{Cao:2023ohj, Li:2024vgv}, 
\begin{equation}
\mathcal T^{\alpha\beta} = (\mathcal E + \mathcal P) u^\alpha u^\beta - \mathcal P g^{\alpha\beta} + \Pi^{\alpha\beta}.
\end{equation}
Here, $u^\alpha \equiv P^\alpha/\sqrt{P^2}$ is the 4-velocity vector. 
The energy density can be extracted as,
\begin{equation}
\mathcal E(r_\perp) =  \int \frac{\dd^2q_\perp}{(2\pi)^2} e^{-i\vec q_\perp \cdot \vec r_\perp} E(q^2_\perp), 
\end{equation}
where, the energy form factor $E(q^2_\perp)$ is defined as, 
\begin{equation}
E(q^2_\perp) = \Big(M+\frac{q^2_\perp}{4M}\Big)A(q^2_\perp) +  \frac{q^2_\perp}{4M} D(q^2_\perp)\,.
\end{equation}
For spin-$\frac{1}{2}$ particles, the energy form factor should include a spin contribution, 
\begin{multline}
E(q^2_\perp) = \Big(M+\frac{q^2_\perp}{4M}\Big)A(q^2_\perp)   \\
+ \frac{q^2_\perp}{4M} \Big[D(q^2_\perp) - 2J(q^2_\perp) \Big]\,.
\end{multline}
This is in contrast to $M^2(q^2_\perp)$, the light-front energy form factor, in which, the spin contribution is absent. 
The energy-momentum conservation (\ref{eqn:A_forward}) \& (\ref{eqn:M2_forward}) [or (\ref{eqn:q2D_forward})] implies 
\begin{equation}
E(0) = M \; \Leftrightarrow \; \int \dd^2 r_\perp \, \mathcal E(r_\perp) = M,
\end{equation}
i.e. energy density is normalized to the total energy in local rest frame, i.e. the mass. 
The 3D r.m.s. radius of the energy distribution is, 
\begin{equation}
r^2_E = \frac{3}{2M}\int \dd^2 r_\perp \, r^2_\perp \mathcal E(r_\perp) = r^2_A - \frac{3}{2} \lambda^2_C \big(1 + D\big).
\end{equation}
The pressure can be extracted as,
\begin{equation}
\mathcal P(r_\perp) = -\frac{1}{6M}\int \frac{\dd^2q_\perp}{(2\pi)^2} e^{-i\vec q_\perp \cdot \vec r_\perp} q^2_\perp D(q_\perp^2).
\end{equation}
The form factor $q^2_\perp D(q^2_\perp)$ is also known as the hadronic cosmological constant \cite{Teryaev:2016edw}, in analogy to the quantum expectation value of the EMT with respect to the vacuum, $\langle 0 | T^{\mu\nu} | 0 \rangle = g^{\mu \nu} \Lambda $. One can readily identify $\Lambda = \frac{1}{2} q^2 D(q^2)$.
The von Laue condition (\ref{eqn:q2D_forward}) implies that hadrons are in mechanical equilibrium, 
\begin{equation}
\int \dd^2 r_\perp \,  \mathcal P(r_\perp) = 0.
\end{equation}
The shear is also related to the GFF $D(q^2_\perp)$, 
\begin{align}
\Pi^{ij}(\vec r_\perp) =\,& \frac{1}{4M}\int \frac{\dd^2q_\perp}{(2\pi)^2} e^{-i\vec q_\perp \cdot \vec r_\perp} \\
& \times \big(q^{i}_\perp q^{j}_\perp - \frac{1}{3}\delta^{ij}q^2_\perp\big) D(q^2_\perp), \\
=\,& \frac{1}{2}\big(\delta^{ij} - 3\hat r^i_\perp \hat r^j_\perp\big) \mathcal P(r_\perp), 
\end{align}
where, $\hat r_\perp^i = r^i_\perp/r_\perp$. The shear is ``traceless", $\Delta_{\alpha\beta}\Pi^{\alpha\beta} = 0$, where $\Delta^{\alpha\beta} = u^\alpha u^\beta - g^{\alpha\beta}$ is the spatial metric tensor. 
In fact, pressure and shear are the trace and traceless parts of the Cauchy stress tensor $\mathcal C^{\alpha\beta} = \Pi^{\alpha\beta} - \mathcal P \Delta^{\alpha\beta}$.

Note that the energy density $\mathcal E(r_\perp)$ is related to the invariant mass squared density $\mathcal M^2(r_\perp)$ through, 
\begin{equation}
\mathcal M^2(r_\perp)  = M \Big[ \mathcal E(r_\perp) - \frac{3}{2} \mathcal P(r_\perp) \Big].
\end{equation}
Physically, the proper energy density $\mathcal E$ is generally assumed positive, a constraint known as the weak energy condition, 
which provides a constraint on $D$:
\begin{equation}\label{eqn:EC}
D \le \frac{2}{3}M^2 r^2_A -1.
\end{equation}
These conditions are trivially satisfied for negative $D$, which is conjectured for stable systems such as the proton \cite{Goeke:2007fp, Perevalova:2016dln}.

A related quantity of interest is the scalar density, $\theta = \mathcal T^{\alpha}_{\;\;\alpha}$, which obeys the classical relation \cite{Rezzolla:2013dea}, 
\begin{equation}
\theta(r_\perp) = \mathcal E (r_\perp) - 3 \mathcal P(r_\perp).
\end{equation}
The scalar radius is $r^2_\theta = r^2_A - \frac{3}{2}\lambda_C^2 (1+ 3D)$.

\begin{figure*}
\centering 
\includegraphics[width=0.5\textwidth]{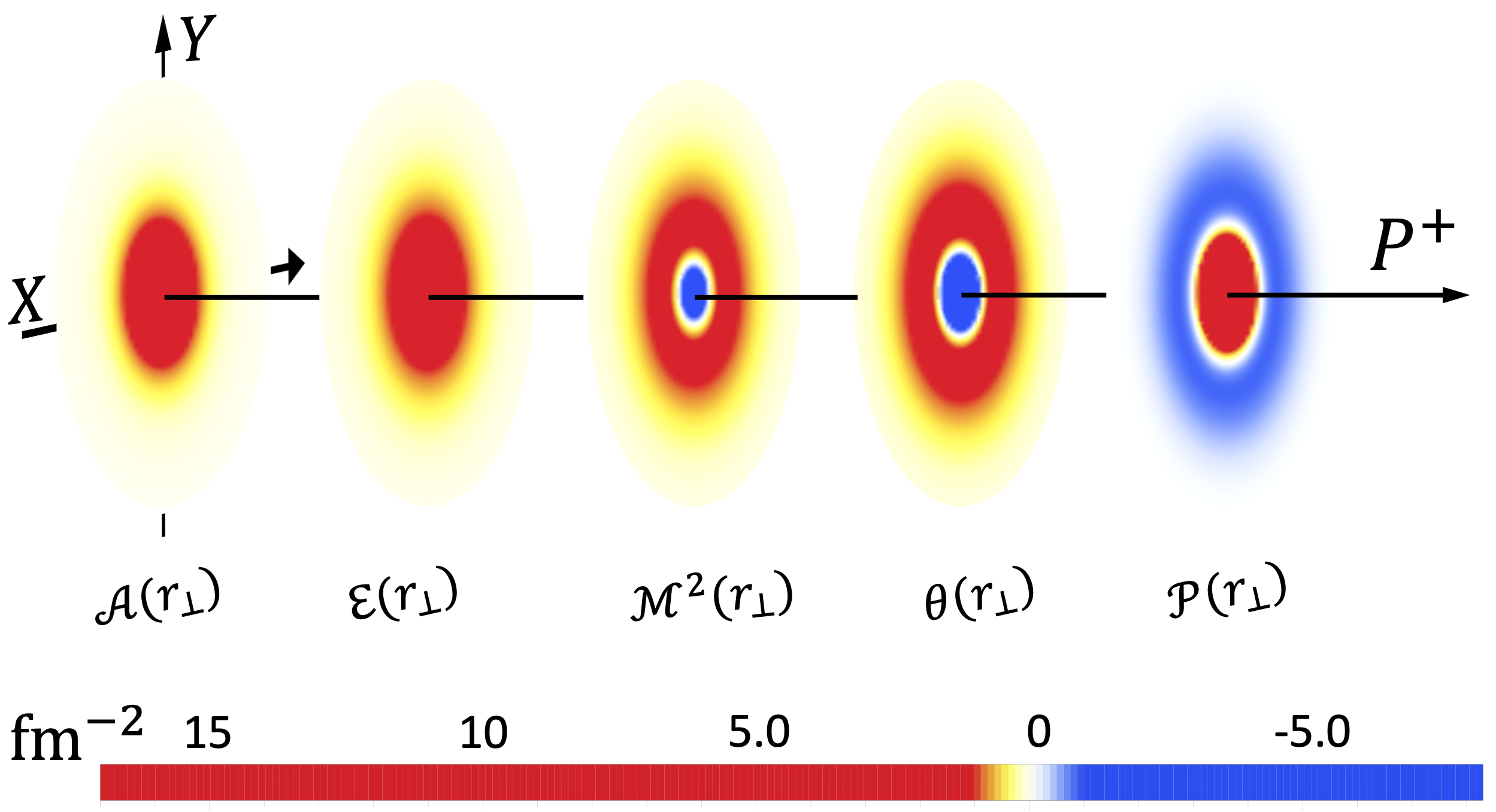}
\caption{A 3D picture of the normalized matter density $\mathcal A(r_\perp)$, energy density $\mathcal E(r_\perp)$, invariant mass squared density $\mathcal M^2(r_\perp)$, aka. internal light-front energy density, scalar density $\theta(r_\perp)$ and pressure $\mathcal P(r_\perp)$ on the light front. All of these densities, except the pressure, are normalized to unity upon integration over the transverse plane. For the pressure, we plot $\mathcal{P}(r_\perp)/M$.}
\label{fig:hadron_radii}
\end{figure*}

The $D$-term is related to the second moment of the pressure, 
\begin{equation}
 D \equiv D(0) =  \frac{3}{2}M \int \dd^2 r_\perp \, r^2_\perp  \mathcal P (r_\perp).
\end{equation}
It is conjectured that $D < 0$ for a mechanically stable system \cite{Goeke:2007fp, Perevalova:2016dln}. Intuitively, a negative $D$ represents a repulsive core along with an attractive periphery. A negative $D$-term suggests a chain of inequalities, 
\begin{equation}
r_\theta > r_{M^2} > r_E. 
\end{equation}
For charmonium, as we will show below, $D \sim -5$. Then, $r_E > r_A$ for this system. Then, we obtain a layered picture of charmonium as shown in Fig.~\ref{fig:hadron_radii}. 

The layered picture is consistent with the physics of QCD. 
From the LFWF representation in Sect.~\ref{sect:LFWF_representation}, partons with large-$x$ contribute most to the matter density $\mathcal A(r_\perp)$. 
Therefore, at the core of a charmonium are the valence quarks. 
For the invariant mass squared density $\mathcal M^2(r_\perp)$, small-$x$ partons, known as ``wee partons", also have a significant contribution since the light-front energy $p^- \propto (p^2_\perp + m^2)/xP^+$. The fact that $r_{M^2} > r_A$ implies that the distribution of the wee partons is broader than the valence quarks. 
The scalar trace $T^\alpha_{\;\;\alpha}$ contains the anomalous gluon fields, whose radius is largest. Therefore, at the outermost of a charmonium, there are glueballs and meson clouds. 
Note that the quark and the antiquark in our model are effective degrees of freedom. Even though there is no explicit gluons, some effects of them have been incorporated. 

In the literature, it is popular to consider the decomposition of the hadron mass into contributions from different species (quarks \& gluon), with each contribution defined from a gauge-invariant operator. Since mass in relativity is not additive, what actually gets decomposed is the proper energy $E$. To remove the kinetic energy contributions, each species is assumed to be stationary. Physically, this decomposition is in analogy to the hydrostatics of coupled multifluids \cite{Rezzolla:2013dea, Lorce:2017xzd}. In principle, the light-front energy can be decomposed in a similar fashion, except that there is no need to assume that the hadron or its components are at rest since the total internal light-front energy $M^2$ is a Lorentz scalar. In the previous work \cite{Cao:2023ohj}, some of us proposed to decompose the light-front energy into a free (or kinetic) part and an interaction (or potential energy) part. This decomposition is more familiar in classical and quantum mechanics although it is not gauge invariant as is the case of the Coulomb potential. Nevertheless, it provides insights into the strong force within charmonium, as we will show in Sect.~\ref{sect:results}.

\section{Light-front wave function representation} \label{sect:LFWF_representation}

\subsection{Gravitational form factor $A$}

In principle, the GFFs $A(-q^2)$ and $D(-q^2)$ can be extracted from any two independent HMEs $t^{\alpha\beta} =  \langle P+\frac{1}{2}q|T^{\alpha\beta}(0)|P - \frac{1}{2}q\rangle/(2P^+)$. We adopt the energy-momentum density $T^{+\mu}$ to extract these form factors. These components provide the correct value of the GFFs in the forward limit as long as the 4-momentum is conserved \cite{Cao:2023ohj}. Furthermore, covariant light-front analysis shows that the hadronic matrix elements of these components are free of spurious zero-mode contributions \cite{Cao:2023ohj}. In the Drell-Yan frame $q^+=0$, 
\begin{align}
t^{++}(\vec q_\perp; P) =\,& P^+ A(q^2_\perp), \\
t^{+i}(\vec q_\perp; P) =\,& P^i_\perp A(q^2_\perp), \\
t^{+-}(\vec q_\perp; P) =\,& \frac{P^2_\perp A(q^2_\perp) + M^2(q^2_\perp)}{P^+},
\end{align}
where, recall, $M^2(q^2_\perp) = (M^2+ \frac{1}{4}q^2_\perp) A(q^2_\perp)  + \frac{1}{2}q_\perp^2 D(q^2_\perp)$.

The GFF $A$ can be extracted from either $T^{++}$ or $T^{+i}$. Indeed, the former is the traditional ``good current". Since $T^{++}$ and $T^{+i}$ do not contain the interaction, $A(Q^2)$ can be represented in terms of the LFWFs as \cite{Brodsky:2000ii},
\begin{multline}\label{eqn:off-forward_++}
A(q^2_\perp) =  \sum_n \sum_{\{s_i\}} \int \big[\dd x_i \dd^2 k_{i\perp} \big]_n \\
\times \sum_j x_j \psi_n(\{x_i, \vec k_{i\perp}, s_i\}) 
\psi_n(\{x_i, \vec k_{i,j\perp}, s_i\}), 
\end{multline}
where, 
\begin{equation}
\vec k_{i,j\perp} = \begin{cases}
\vec k_{i\perp} - x_i \vec q_\perp, &  \text{spectator: }i\ne j \\
\vec k_{i\perp} + (1- x_i) \vec q_\perp, & \text{struck parton: } i = j\\
\end{cases}
\end{equation}
And the $n$-body integration measure is defined as, 
\begin{multline}
\int\big[\dd x_i \dd^2 k_{i\perp}\big]_n = \frac{1}{S_n}\prod_{i=1}^{n} \int \frac{\dd x_i}{2x_i} 2\delta(\sum_i x_i -1) \\
\times \int \frac{\dd^2k_{i\perp}}{(2\pi)^3} (2\pi)^3\delta^2(\sum_i k_{i\perp})
\end{multline}
where, $S_n$ is the symmetry factor. 
Using the coordinate space representation developed in our previous work \cite{Cao:2023ohj}, $A(q^2_\perp)$ becomes, 
\begin{multline}\label{eqn:A_coord}
A(q_\perp^2) =\sum_n \sum_{\{s_i\}} \int \big[\dd x_i \dd^2 r_{i\perp}\big]_n  \\
\times \Big| \widetilde\psi_n(\{x_i, \vec r_{i\perp}, s_i\})\Big|^2 \sum_j x_j e^{i \vec r_{j\perp}\cdot \vec q_\perp}\,,
\end{multline}
where, 
\begin{multline}
\int\big[\dd x_i \dd^2 r_{i\perp}\big]_n = \frac{1}{S_n}\prod_{i=1}^{n} \int \frac{\dd x_i}{4\pi x_i} 4\pi \delta(\sum_i x_i -1) \\
\times \int \dd^2r_{i\perp} \delta^2(\sum_i x_i\vec r_{i\perp})\,.
\end{multline}
Then, the matter distribution becomes a one-body density (OBD), 
\begin{multline}\label{eqn:density_A}
\mathcal A(r_\perp) 
= \sum_n \sum_{\{s_i\}} \int \big[\dd x_i \dd^2 r_{i\perp}\big]_n \Big| \widetilde\psi_n(\{x_i, \vec r_{i\perp}, s_i\})\Big|^2 \\
\times \sum_j x_j \delta^2(\vec r_\perp - \vec r_{j\perp})\,.
\end{multline}
And the corresponding r.m.s. matter radius is,
\begin{multline}
r^2_A \equiv -6 \frac{1}{A(0)}\frac{\dd A(Q^2)}{\dd Q^2 }\Big|_{Q^2\to 0}, \\
= \frac{3}{2} \sum_n \sum_{\{s_i\}} \int \big[\dd x_i \dd^2 r_{i\perp}\big]_n   
 \Big| \widetilde\psi_n(\{x_i, \vec r_{i\perp}, s_i\})\Big|^2 \\
\times \sum_j x_j r_{j\perp}^2.
\end{multline}
As we mentioned, the matter density mainly samples contributions from the large-$x$ partons, i.e. valence quarks. 

In this work, we only consider the valence Fock sector contribution. The LFWF representation in this ansatz can be written as,
\begin{multline}
A(q_\perp^2) = \sum_{s, \bar s}\int_0^1\frac{\dd x}{2x(1-x)} \int\frac{\dd^2k_\perp}{(2\pi)^3} 
\psi_{s\bar s}(x, \vec k_\perp)  \Big\{  (1-x) \\
\times \psi^*_{s\bar s}(x, \vec k_\perp-x\vec q_\perp) 
 + x \psi^*_{s\bar s}(x, \vec k_\perp+(1-x)\vec q_\perp)
\Big\} , 
\end{multline}
Similarly, the matter density reads, 
\begin{multline}
\mathcal A(\vec r_\perp) 
= \sum_{s, \bar s}\int_0^1\frac{\dd x}{4\pi x(1-x)} 
\Big\{ (1-x) \Big| \widetilde \psi_{s\bar s}\Big(x, \frac{\vec r_\perp}{x}\Big) \Big|^2 \\
 + x \Big| \widetilde \psi_{s\bar s}\Big(x, -\frac{\vec r_\perp}{1-x}\Big) \Big|^2 
\Big\}. 
\end{multline}

In basis representation, 
\begin{multline}\label{eqn:A_basis_representation}
A(q^2_\perp) = \sum_{s, \bar s, n, l, n', l'} \psi^*_{s\bar s}(n', m', l')\psi_{s\bar s}(n, m, l)  \int_0^1 \frac{\dd x}{4\pi}   \\
 \times  \chi_{l'}(x) \chi_{l}(x)\int \frac{\dd^2 k_\perp}{(2\pi)^2} \Big[ 
  (1-x) \phi^*_{n'm'}\big(\vec k_\perp-\sqrt{\frac{x}{1-x}}\vec q_\perp\big) \\
 +x \phi^*_{n'm'}\big(\vec k_\perp+\sqrt{\frac{1-x}{x}}\vec q_\perp\big)\Big] \phi_{nm}(\vec k_\perp) \,.
\end{multline}
Note that $m= m' = m_j - s - \bar s$ according to angular momentum conservation, but is otherwise arbitrary though limited by the total angular momentum of the state. 
Similarly, the basis representation of the matter density $\mathcal A(\vec r_\perp)$ reads, 
\begin{multline}
\mathcal A(\vec r_\perp) 
=  \sum_{s, \bar s, n, l, n', l'} \psi^*_{s\bar s}(n', m', l')\psi_{s\bar s}(n, m, l)  \int_0^1 \frac{\dd x}{4\pi}  \\
 \times  \chi_{l'}(x) \chi_{l}(x)
 \Big[
\frac{x^2}{1-x} \widetilde\phi^*_{n'm'}\Big(\sqrt{\frac{x}{1-x}} \vec r_\perp\Big) \widetilde\phi_{nm}\Big( \sqrt{\frac{x}{1-x}} \vec r_\perp\Big) \\
+ 
\frac{(1-x)^2}{x} \widetilde\phi^*_{n'm'}\Big(\sqrt{\frac{1-x}{x}} \vec r_\perp\Big) \widetilde\phi_{nm}\Big( \sqrt{\frac{1-x}{x}} \vec r_\perp\Big)\Big].
\end{multline}

\subsection{Gravitational form factor $D$: kinetic part}

The GFF $D(q^2_\perp)$ and the light-front energy form factor $M^2(q^2_\perp)$ can be extracted from either the light-front energy density $T^{+-}$, or the transverse stress $T^{11}+T^{22}$ and $T^{12}$. In practical model calculations, such as this work, truncations and approximations are introduced which result in the loss of some Poincaré symmetries. One then deals with a system with a reduced number of symmetries, and GFFs extracted from different current components are no longer equivalent. One possible way to extract the form factors is to parametrize the HMEs in terms of the reduced symmetry. Effectively, one introduces additional Lorentz tensor structures. Form factors associated with non-covariant structures are called spurious ones. See Ref.~{\cite{Cao:2023ohj}} and the references therein for a discussion on this approach to the EMT. From the covariant light-front dynamics analysis \cite{Cao:2023ohj}, $T^{11}+T^{22}$  and $T^{12}$ are contaminated by spurious form factors and may violate the von Laue condition\footnote{We recently discovered a new symmetry that may protect $T^{12}$ from the contamination of the spurious form factor \cite{Cao:2024rul}. It is not clear whether this symmetry is applicable to the present system.}. $T^{+-}$ is the light-front energy density and it contains the interaction. While this operator is more involved, it is constrained by energy conservation which ensures the von Laue condition even after model truncation. 
We therefore use $T^{+-}$ to extract GFF $D$. Our method is general enough and can be applied to other bound-state systems. 
The HME of $T^{+-}$ consists of two parts: 
\begin{equation}
t^{+-} = \frac{P^2_\perp A(q^2_\perp) + M^2(q^2_\perp)}{P^+}\,.
\end{equation}
In the Breit frame $P_\perp = 0$, only the light-front energy form factor $M^2(q^2_\perp)$ survives. Form factor $M^2(q^2_\perp)$ is related to GFFs $A$ and $D$ as given in (\ref{eqn:M2FF}).

In Ref.~\cite{Cao:2023ohj}, we propose to split $T^{+-}$ into two parts $T^{+-} = T^{+-}_0 + T^{+-}_\text{int}$: a free part $T^{+-}_0$ and an interaction part $T^{+-}_\text{int}$. We adopt the light-cone gauge $A^+=0$ and in this gauge, the kinetic part $T^{+-}_\text{0}$ is simple. The corresponding invariant mass squared form factor is also split into two pieces, $M^2(q^2_\perp) = M^2_0(q^2_\perp) + M^2_\text{int}(q^2_\perp)$.
The free energy density $T^{+-}_0$ is diagonal in Fock space, and the corresponding invariant mass squared form factor admits an exact LFWF representation,  
\begin{multline}
M^2_0(q^2_\perp) = \int \big[\dd x_i \dd^2 k_{i\perp} \big]_n \sum_j \psi_n^*(\{x_i, \vec k_{i,j\perp}^+, s_i\})
 \\
\times \frac{\vec k_{j\perp}^2+m_j^2-\frac{1}{4}q_\perp^2}{x_j} \psi_n(\{x_i, \vec k_{i,j\perp}^-, s_i\}), 
\end{multline}
where, 
\begin{align*}
\vec k^+_{i,j\perp} =\,& \begin{cases}
\vec k_{i\perp} + \frac{1}{2} x_i \vec q_\perp, &  \text{spectator: }i\ne j \\
\vec k_{i\perp} - \frac{1}{2}(1- x_i) \vec q_\perp, & \text{struck parton: } i = j\\
\end{cases} \\
\vec k^-_{i,j\perp} =\,& \begin{cases}
\vec k_{i\perp} - \frac{1}{2} x_i \vec q_\perp, &  \text{spectator: }i\ne j \\
\vec k_{i\perp} + \frac{1}{2}(1- x_i) \vec q_\perp, & \text{struck parton: } i = j\\
\end{cases}
\end{align*}
The integrand is the $n$-body kinetic energy with an extra term $-\frac{1}{4}q^2_\perp$. This term accounts for the recoil effect, which has the same origin as the $D$-term as one can see from the free boson case \cite{Pagels:1966zza} (cf. \cite{Cao:2023ohj}). 

Using the coordinate space LFWFs, $M^2_0(q^2_\perp)$ is dramatically simplified as, 
\begin{multline}\label{eqn:M20}
M_0^2(q^2_\perp) = \sum_n \sum_{\{s_i\}} \int \big[\dd x_i \dd^2 r_{i\perp} \big]_n \widetilde\psi_n^*(\{x_i, \vec r_{i\perp}, s_i\}) \\
\times \sum_j  e^{i\vec r_{j\perp}\cdot\vec q_\perp}  \frac{-\vec\nabla_{j\perp}^2+m_j^2-\frac{1}{4}q_\perp^2}{x_j}  
 \widetilde\psi_n(\{x_i, \vec r_{i\perp}, s_i\}) \,.
 \end{multline}
 Its Fourier transform gives the OBD of the free light-front energy, 
\begin{multline}
\mathcal M_0^2(r_\perp) = \sum_n \sum_{\{s_i\}} \int \big[\dd x_i \dd^2 r_{i\perp} \big]_n \sum_j  \delta^2(\vec r_{j\perp} - \vec r_\perp)  
\\
\times \bigg\{
\widetilde\psi_n^*(\{x_i, \vec r_{i\perp}, s_i\}) \frac{-\vec\nabla_{j\perp}^2+m_j^2}{x_j}  
 \widetilde\psi_n(\{x_i, \vec r_{i\perp}, s_i\}) \\
  +\frac{1}{4x_j}\nabla^2_{j\perp} \big| \widetilde\psi_n(\{x_i, \vec r_{i\perp}, s_i\}) \big|^2 \bigg\}\,.
 \end{multline}
For charmonium, 
\begin{multline}
M^2_0(q_\perp^2) = \sum_{s, \bar s}\int \frac{\dd x}{2x(1-x)} \int \frac{\dd^2k_\perp}{(2\pi)^3} \Big\{ \\
  \psi^*_{s\bar s}(x, \vec k_\perp-\half (1-x)\vec q_\perp) 
   \frac{k_\perp^2+m_{q}^2-\frac{1}{4}\vec q_\perp^2}{x} \\
 \times \psi_{s\bar s}(x, \vec k_\perp+\half (1-x)\vec q_\perp) + \psi^*_{s\bar s}(x, \vec k_\perp+\half x\vec q_\perp) \\
 \times \frac{k_\perp^2+m_{\bar q}^2-\frac{1}{4}\vec q_\perp^2}{1-x} 
 \psi_{s\bar s}(x, \vec k_\perp-\half x\vec q_\perp)
 \Big\} \,.
\end{multline}
In basis representation, 
\begin{multline}
M^2_0(q_\perp^2) =  \sum_{s, \bar s, n, l, n', l'} \psi^*_{s\bar s}(n', m', l')\psi_{s\bar s}(n, m, l) \int_0^1 \frac{\dd x}{4\pi}  \\
  \times \chi_{l'}(x) \chi_{l}(x)  \Big\{
\int\frac{\dd^2 k_\perp}{(2\pi)^2}  \phi^*_{n'm'} \big(\vec k_\perp-\frac{1}{2} \sqrt{\frac{1-x}{x}}\vec q_\perp\big) \\
\times \big[ (1-x)k_\perp^2+\frac{m_q^2-\frac{1}{4}q_\perp^2}{x} \big] \phi_{nm} \big(\vec k_\perp+\frac{1}{2} \sqrt{\frac{1-x}{x}}\vec q_\perp\big)  \\
 + \phi^*_{n'm'}\big(\vec k_\perp+\half\sqrt{\frac{x}{1-x}}\vec q_\perp\big) \big[ x k_\perp^2+\frac{m_{\bar q}^2-\frac{1}{4}q_\perp^2}{1-x}\big] \\
 \times \phi_{nm}\big(\vec k_\perp-\half\sqrt{\frac{x}{1-x}}\vec q_\perp\big)
 \Big\} \,.
\end{multline}

\subsection{Gravitational form factor $D$: interaction part}

The interaction part of the EMT in coordinate space can be formally represented in Fock space using LFWFs. 
However, the expression is not particularly useful in quantum field theories, since it involves particle number changing. 
Furthermore, the interaction $v_{nm}$ also has to be renormalized. The effective interactions, such as the one-gluon-exchange we employed, are non-local in general. It may seem that a local interaction density has to be obtained from an underlying field theory. 
On the other hand, in non-relativistic quantum many-body theory, the Hamiltonian density can also be obtained by localizing the Hamiltonian operator, viz. inserting $\delta^3(r - r_i)$ for the $i$-th constituent. For example, the charge density can be obtained by localizing the charge,
\begin{equation}
Q  = \sum_i e_i \quad \to \quad \rho(r) = \sum_i e_i \delta^3(r - r_i)\,.
\end{equation} 
The above method can be generalized to the relativistic quantum many-body systems. 
In relativistic quantum theory, particles cannot be consistently localized\footnote{For mildly relativistic systems ($v/c \sim 0.1$), Newton-Wigner operator can be used to localize particles \cite{Pryce:1948pf}. The application of this operator to QCD can be found in Ref.~\cite{Lorce:2018zpf}.}. Fortunately, there is one way to circumvent this problem using light-front dynamics. In light-front dynamics, particles can be localized on the transverse plane using the Dirac-$\delta$ prescription $\delta^2(r_\perp - r_{i\perp})$, which is sufficient to extract the transverse densities. For example, the transverse charge density can be obtained by localizing the charge on the transverse plane, 
\begin{equation}
Q  = \sum_i e_i \quad \to \quad \rho(r_\perp) = \sum_i e_i \delta^2(r_\perp - r_{i\perp})\,.
\end{equation} 
One can show that the charge density obtained in this way reproduces the Drell-Yan-West formula \cite{Soper:1976jc, Brodsky:2006uqa}. Similarly, the matter density can also be obtained using particle localization on the light front. 

The effective interaction we employed is a two-body potential,
\begin{equation}
V = \frac{1}{2!}\sum_{i, j} v(x, r_{i\perp} - r_{j\perp}),
\end{equation}
where, $\vec r_{i\perp}$ is the transverse coordinate of the $i$-th parton. Based on the localization prescription, we can write down the corresponding energy density as, 
\begin{multline}
\mathcal V(r_\perp) = \frac{1}{2!}\sum_{i, j} v(x, r_{i\perp} - r_{j\perp}) \\
\times \frac{1}{2}\big\{ \delta^2(r_\perp - r_{i\perp}) + \delta^2(r_\perp - r_{j\perp}) \big\}\,.
\end{multline}
Here, the inter-particle energy density is evenly distributed between the two interacting constituents. The above many-body expression can be converted into the corresponding operator following the standard second quantization rule. 

To obtain the inter-particle energy $v(x, r_{i\perp} - r_{j\perp})$, we first observe that the interaction energy in the forward limit admits a  diagonal representation, 
\begin{align}
M^2_\text{int}(0) =\,& \sum_n \sum_{\{s_i\}} \int \big[\dd x_i \dd^2 k_{i\perp} \big]_n \psi_n^*(\{x_i, \vec k_{i\perp}, s_i\}) \\
 & \times \psi_n(\{x_i, \vec k_{i\perp}, s_i\})  v_n(\{x_i, \vec k_{i\perp}\}), \\
  =\,& \sum_n\sum_{\{s_i\}} \int \big[\dd x_i \dd^2 r_{i\perp} \big]_n \widetilde\psi_n^*(\{x_i, \vec r_{i\perp}, s_i\}) \\
  & \times v_n(\{x_i, -i\vec \nabla_{i\perp}\})
\widetilde\psi_n(\{x_i, \vec r_{i\perp}, s_i\}) ,
\end{align}
where, the $n$-body potential energy is expressed in terms of the mass eigenvalue and the kinetic energy 
\begin{equation}
v_n(\{x_i, \vec k_{i\perp}\}) = M^2 - \sum_{j=1}^n \frac{\vec k_{j\perp}^2+m_j^2}{x_j}.
\end{equation}
 This representation is exact and can be obtained directly from the Schrödinger equation (\ref{eqn:ev_P}). 

To generalize the above expression to the off-forward case, we need to specify the locations of the interactions. We adopt the impulse ansatz that all interactions happen at the same instant in light-front time. This ansatz is certainly held for local interactions. In Ref.~\cite{Cao:2023ohj}, we rigorously showed that this ansatz is valid in a pion cloud model where all scattering appears at the location of the nucleon.  For small-size systems, such as charmonium, this ansatz is expected to be a good approximation. For effective interactions, this ansatz requires neglecting the propagation of the intermediate particles, which is widely adopted in non-relativistic systems. Since we do not incorporate dynamical gluons and sea quarks in this work, this ansatz is consistent with our approximation. 

The interaction form factor in this ansatz becomes,
\begin{multline}
 M^2_\text{int}(q^2_\perp) 
  = \sum_n \frac{1}{n}\sum_{\{s_i\}} \int \big[\dd x_i \dd^2 r_{i\perp} \big]_n \widetilde\psi_n^*(\{x_i, \vec r_{i\perp}, s_i\}) \\
\times   \sum_{j} e^{i\vec q_\perp \cdot \vec r_{j\perp}} v_n(\{x_i, -i\vec \nabla_{i\perp}\})
\widetilde\psi_n(\{x_i, \vec r_{i\perp}, s_i\}),
\end{multline}
assuming the interaction is two-body. 
For charmonium, the above expression becomes, 
\begin{multline}
M^2_\text{int}(q^2_\perp) =  \frac{1}{2}\sum_{s, \bar s}\int\frac{\dd x}{4\pi x(1-x)} \int \dd^2r_\perp 
\widetilde\psi^*_{s\bar s}(x, \vec r_\perp) \\
\times \Big[ e^{i\vec q_\perp \cdot \vec r_{1\perp}} + e^{i\vec q_\perp \cdot \vec r_{2\perp}} \Big] v(x, \vec r_\perp, -i\vec\nabla_\perp) \widetilde\psi_{s\bar s}(x, \vec r_\perp) ,
\end{multline}
where, $\vec r_{1\perp} = \vec R_\perp + (1-x)\vec r_\perp =  (1-x)\vec r_\perp$, $r_{2\perp} = \vec R_\perp - x_1 \vec r_\perp = -x\vec r_\perp$, and 
 \begin{equation}
 v(x, \vec r_\perp, -i\vec\nabla_\perp) = M^2 - \frac{- \nabla_\perp^2+m_q^2}{x} - \frac{- \nabla_\perp^2+m^2_{\bar q}}{1-x}.
 \end{equation}
In terms of the momentum-space LFWFs, the above expression can also be written as, 
 \begin{multline}
M^2_\text{int}(q^2_\perp) = \frac{1}{2} \sum_{s, \bar s}\int\frac{\dd x}{2 x(1-x)} \int \frac{\dd^2k_\perp}{(2\pi)^3} \\
\times \Big[ \psi^*_{s\bar s}\big(x, \vec k_\perp+(1-x)\vec q_\perp\big)  
+ \psi^*_{s\bar s}(x, \vec k_\perp-x\vec q_\perp)  \Big]\\
\times \Big[ M^2 - \frac{k_\perp^2+m_q^2}{x} - \frac{k_\perp^2+m^2_{\bar q}}{1-x} \Big] \psi_{s\bar s}(x, \vec k_\perp) \,.
\end{multline}
In basis representation,
\begin{multline}
M^2_\text{int}(q^2_\perp) 
 =  \frac{1}{2} \sum_{s, \bar s, n, l, n', l'} \psi^*_{s\bar s}(n', m', l')\psi_{s\bar s}(n, m, l) \int_0^1 \frac{\dd x}{4\pi} \\
 \times   \chi_{l'}(x) \chi_{l}(x)  \int\frac{\dd^2 k_\perp}{(2\pi)^2}   \phi_{nm}\big(\vec k_\perp\big)  \Big( M^2 - k_\perp^2 - \frac{m_c^2}{x(1-x)} \Big) \\
\times \Big[ \phi^*_{n'm'} \big(\vec k_\perp+ \sqrt{\frac{1-x}{x}}\vec q_\perp\big) 
 + \phi^*_{n'm'}\big(\vec k_\perp-\sqrt{\frac{x}{1-x}}\vec q_\perp\big) \Big] \,.
 \end{multline}

In the forward limit ($q^2 = 0$), it is evident that
\begin{equation}
M^2_0(0)+M^2_\text{int}(0) = M^2,
\end{equation}
from the LFWF representation. From (\ref{eqn:M2FF}), this condition implies 
\begin{equation}
\lim_{q^2_\perp \to 0} q^2_\perp D(q^2_\perp) = 0.
\end{equation}
Therefore, the forward limit constraint is satisfied. 

\section{Numerical results}\label{sect:results}

We substitute the charmonium LFWFs obtained from BLFQ \cite{Li:2019}. 
As mentioned, the form factors are evaluated with $N_\text{max} = L_\text{max} = 8$. The uncertainty associated with the basis resolution is quoted as the difference between the $N_\text{max} = L_\text{max} = 8$ results and the $N_\text{max} = L_\text{max} = 16$ results.

\begin{figure}
\centering 
\includegraphics[width=0.45\textwidth]{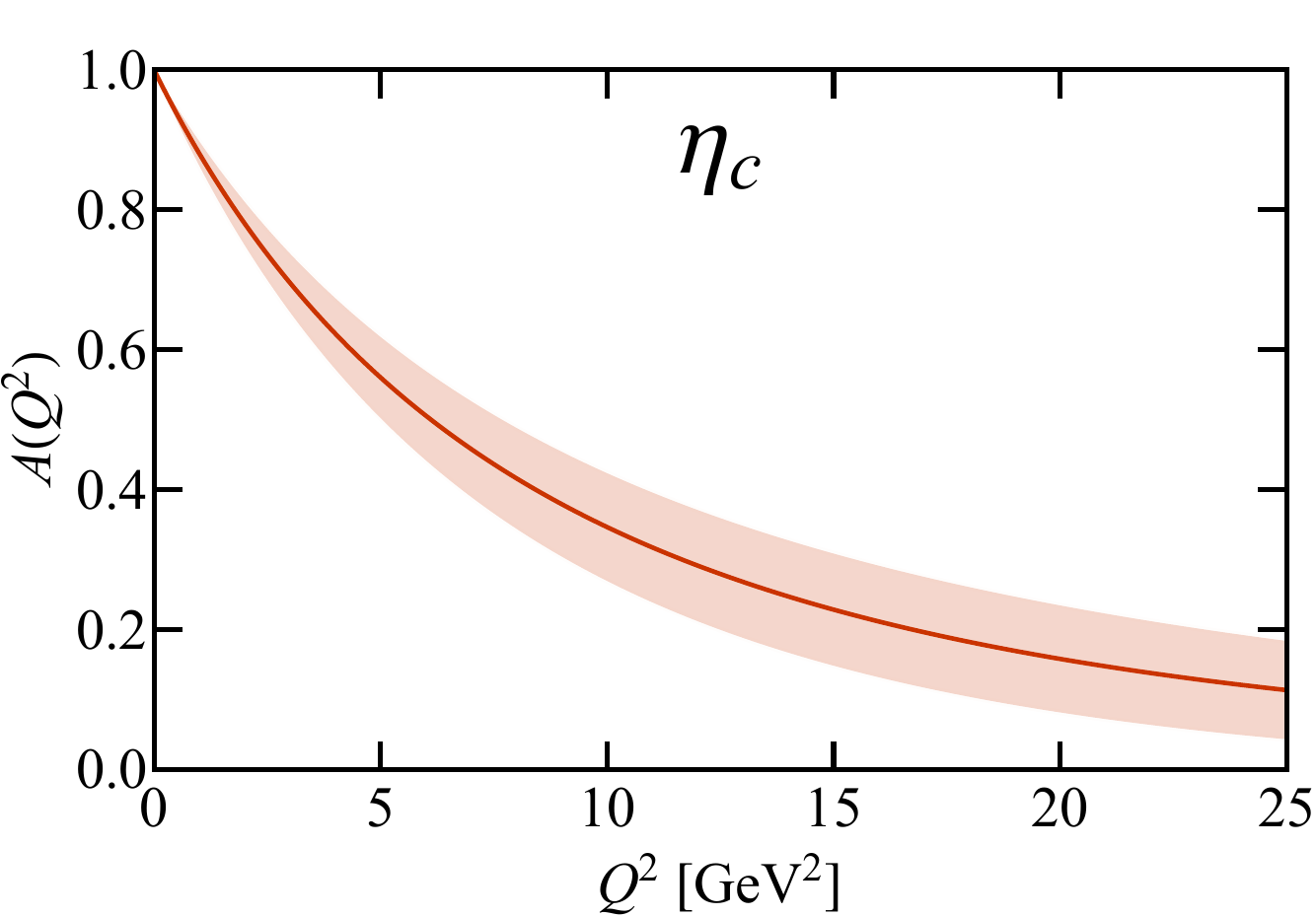}
\includegraphics[width=0.45\textwidth]{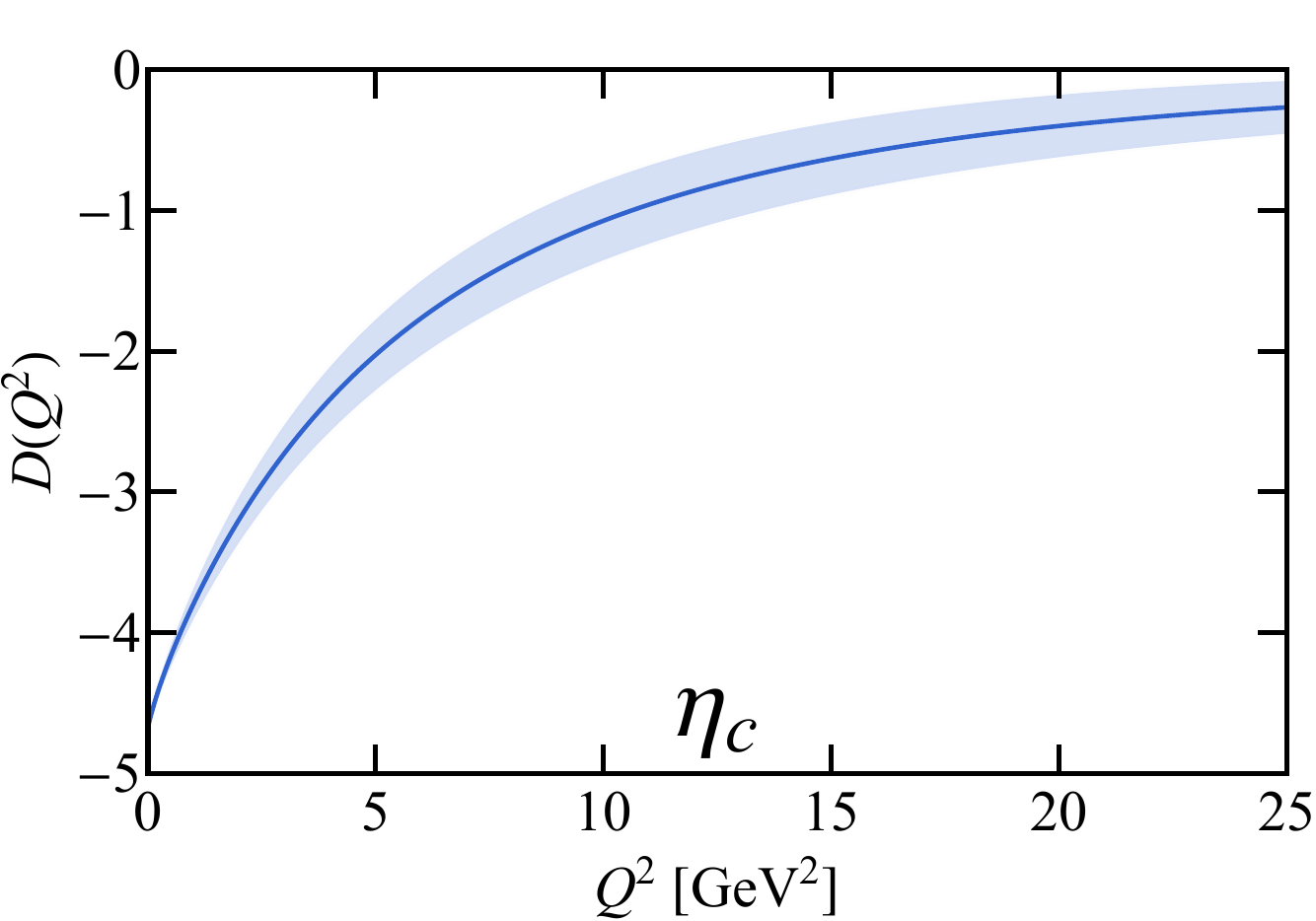}
\caption{GFFs $A(Q^2)$ (\textit{top}) and $D(Q^2)$ (\textit{bottom}) of ground-state pseudoscalar charmonium $\eta_c$. The central value is based on $N_\text{max} = 8$ wave functions and the difference between $N_\text{max} = 8$ and $N_\text{max} = 16$ results is shown as the bands to indicate the basis sensitivity.}
\label{fig:etac_GFF_AD}
\end{figure}

\begin{table*}
\caption{BLFQ prediction of the $D$-term and the radii of charmonium $\eta_c$, $\eta_c'$ and $\chi_{c0}$. We adopt the $N_{\text{max}} = 8$ results as the central values, which corresponds to a UV resolution about the same as the hadron mass. In parentheses we quote the difference between  $N_{\text{max}} = 8$ and  $N_{\text{max}} = 16$ results as the basis sensitivity. }
\label{tab:D_radii}
\begin{tabular}{p{1.5cm}p{1.5cm}p{1.5cm} p{1.5cm}p{1.5cm}p{1.5cm} p{2cm}}
\toprule
 & $D$ & $r_A$ (fm) & $r_E$ (fm) & $r_{M^2}$ (fm) & $r_\theta$ (fm) & ${2\langle T \rangle}/{\langle V \rangle}$ \\
 \colrule
 $\eta_c$ & $-4.7(0)$ &  $0.18(1)$ & $0.23(1)$ & 0.29(1) & 0.34(0) & $-1.7(0)$ \\
 $\eta_c'$ & $-6.7(4)$ & $0.35(2)$ & $0.39(2)$ & 0.42(2) & 0.46(2) & $-14(8)$ \\
 $\chi_{c0}$ & $-6.2(1)$ & $0.24(2)$ & $0.29(1)$ & 0.34(1) & 0.39(1) & $-3.0(4)$ \\
 \botrule
 \end{tabular}
 \end{table*}

Figure~\ref{fig:etac_GFF_AD} shows the GFFs $A(Q^2)$, $D(Q^2)$ for the ground-state pseudoscalar charmonium $\eta_c$. From these results, we can extract  $D$ and radius $r_A$. The results are listed in Table~\ref{tab:D_radii}. Also listed in Table~\ref{tab:D_radii} are the radii, $r_E$, $r_{M^2}$, $r_\theta$, as combinations of these two primary quantities. 
A somewhat surprising result is the large contribution from terms proportional to $\lambda_C = 1/M = 0.066\,\mathrm{fm}$. Naïvely, for quarkonium, this term is expected to be small. However, the matter radius is roughly proportional to the inverse of the binding energy $r_A \sim B^{-1} \sim \alpha^{-1}_s \lambda_C$, where the strong coupling constant $\alpha_s \sim 0.3$ for charmonium. Therefore, $r_A$ and $\lambda_C$ are actually of the same order of magnitude. 

\begin{figure}
\centering 
\includegraphics[width=.45\textwidth]{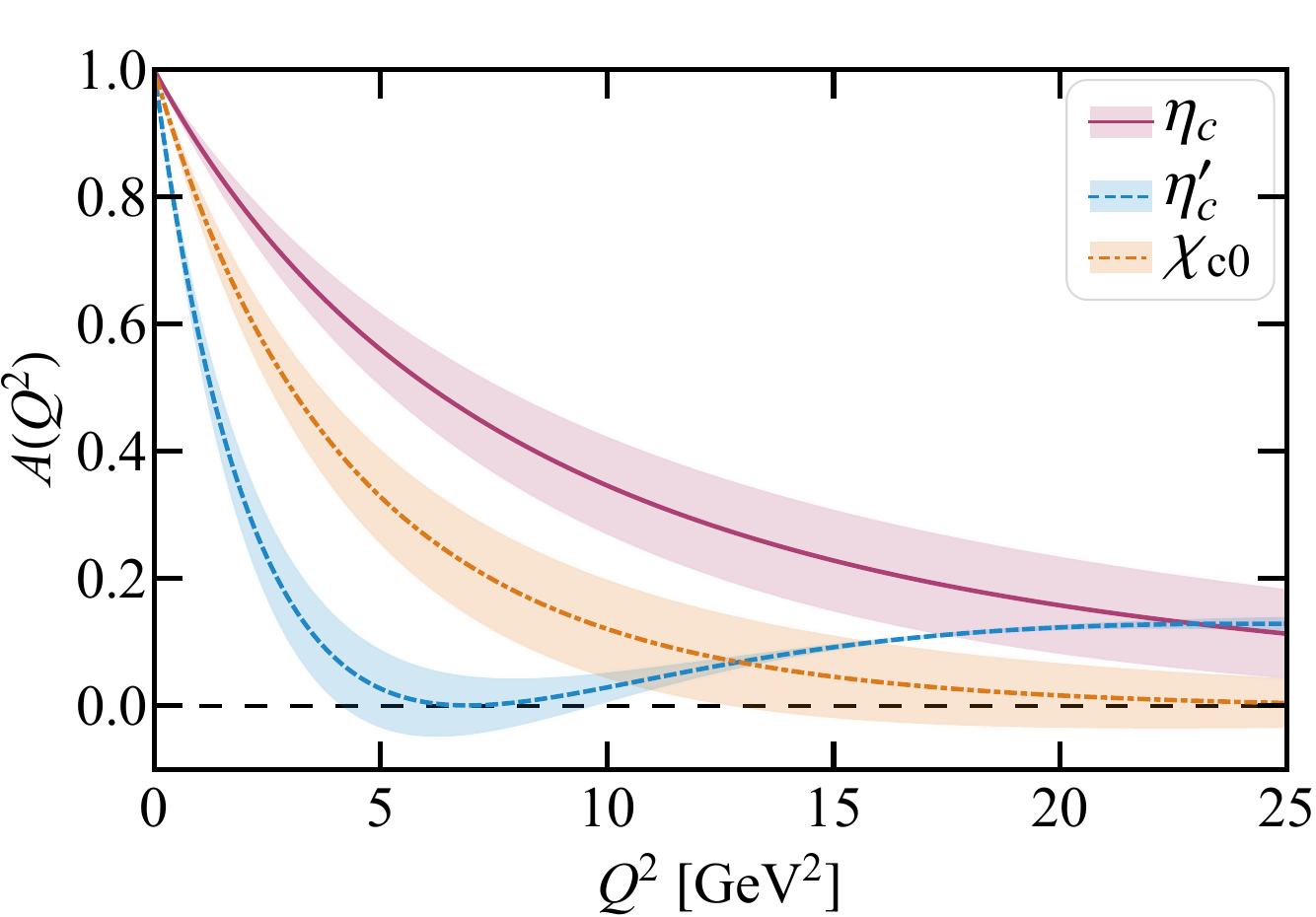}
\includegraphics[width=.45\textwidth]{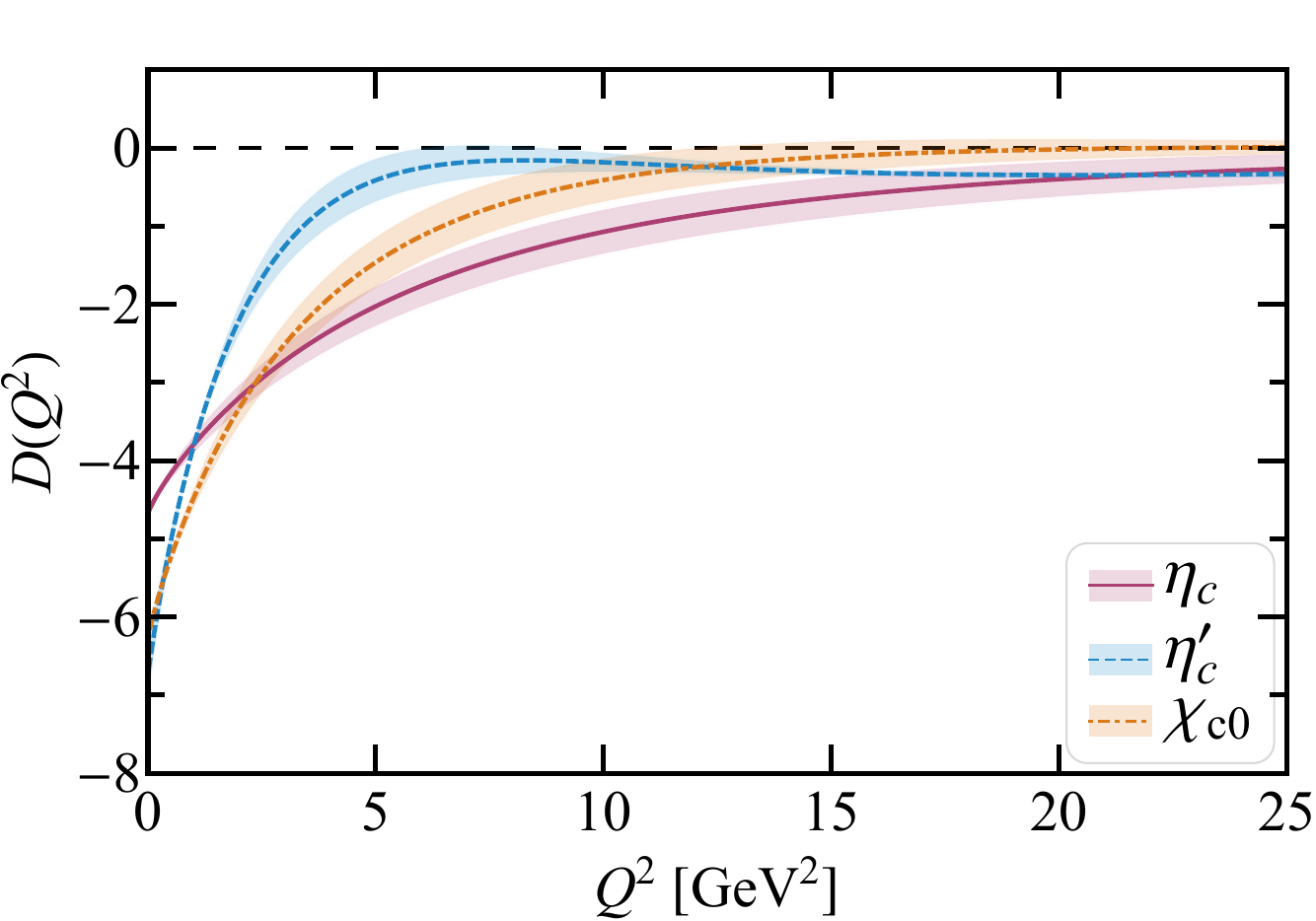}
\caption{Comparison of GFFs $A(Q^2)$ (\textit{top}) and $D(Q^2)$ (\textit{bottom}) of charmonia $\eta_c$, $\eta_c'$ and $\chi_{c0}$. The uncertainty bands are determined as described in the caption to Fig.~\ref{fig:etac_GFF_AD}.}
\label{fig:charmonia_GFF_AD}
\end{figure}

Figure~\ref{fig:charmonia_GFF_AD} compares the GFFs for $\eta_c$ with those for its radial and angular excitations, viz. $\eta'_c$ and $\chi_{c0}$. Radial and angular nodes appear for excited states as expected. The corresponding $D$-term and the radii are extracted and collected in Table~\ref{tab:D_radii}. From these numbers, the matter radius $r_A$ of the $P$-wave charmonium $\chi_{c0}$ is larger than that of $\eta_c$ by 30\%. The radius $r_A$ of the $\eta_c'$, the radial excitation, almost doubles that of the ground state. 

\begin{figure}
\centering 
\includegraphics[width=0.475\textwidth]{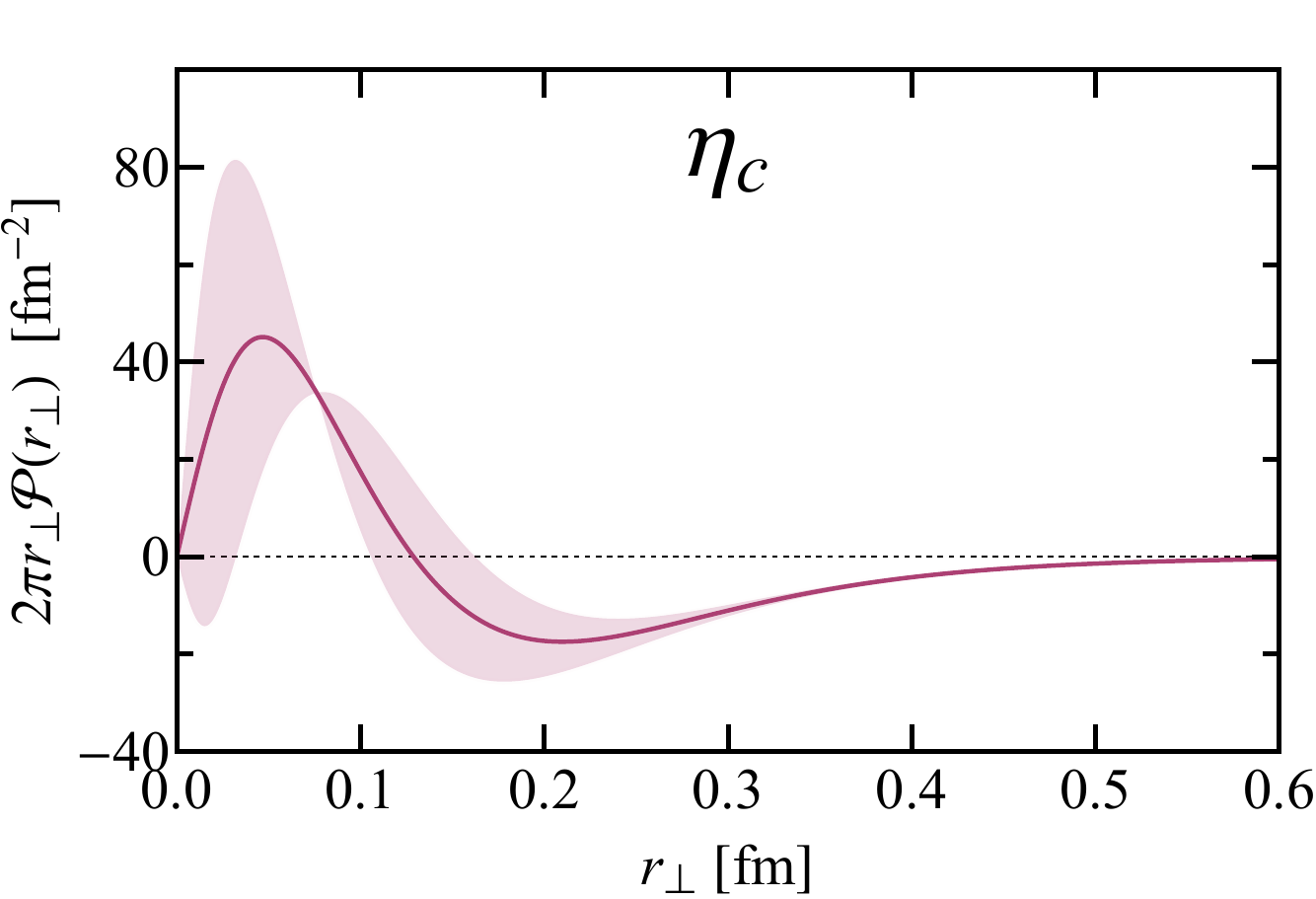}
\includegraphics[width=0.475\textwidth]{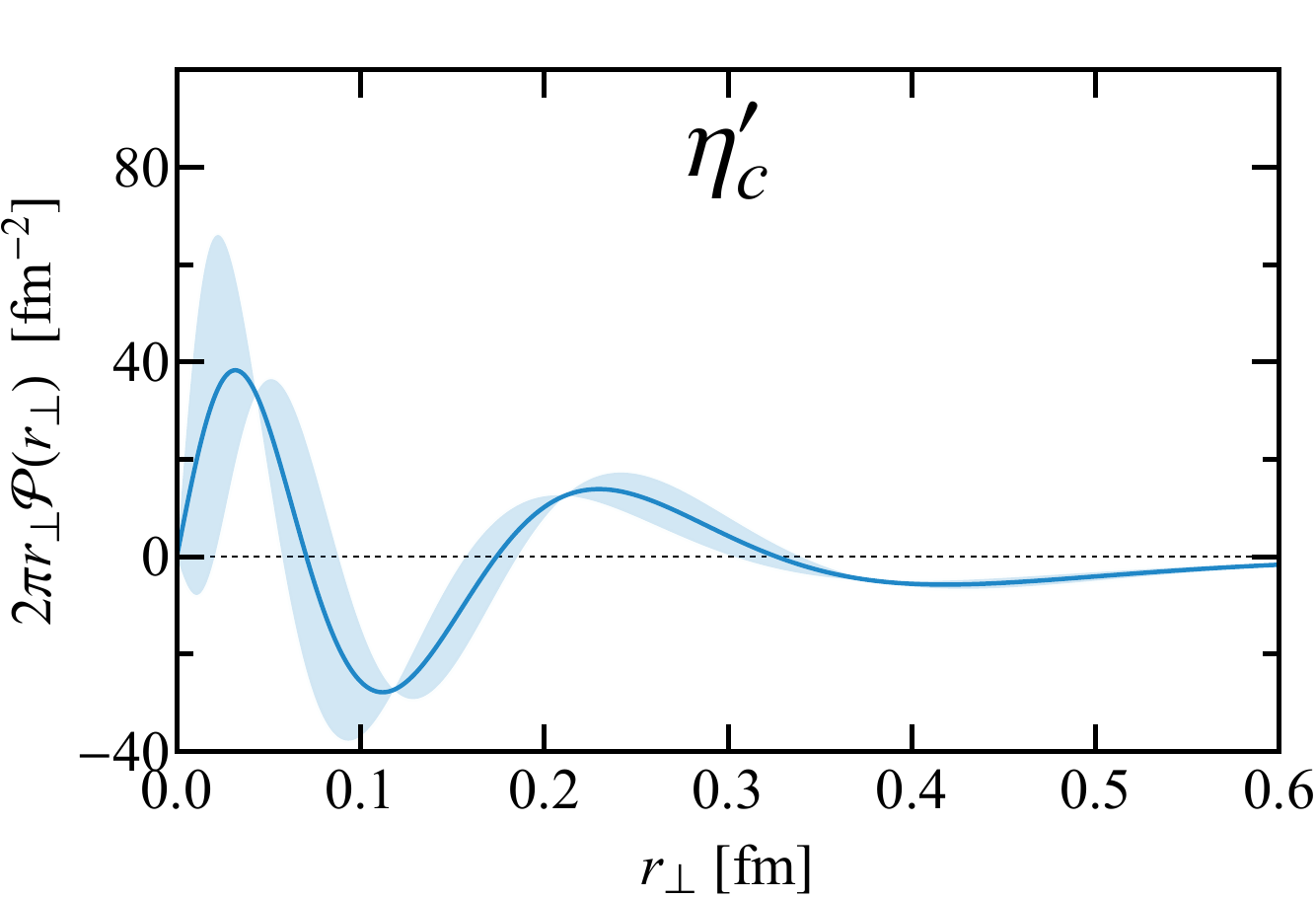}
\includegraphics[width=0.475\textwidth]{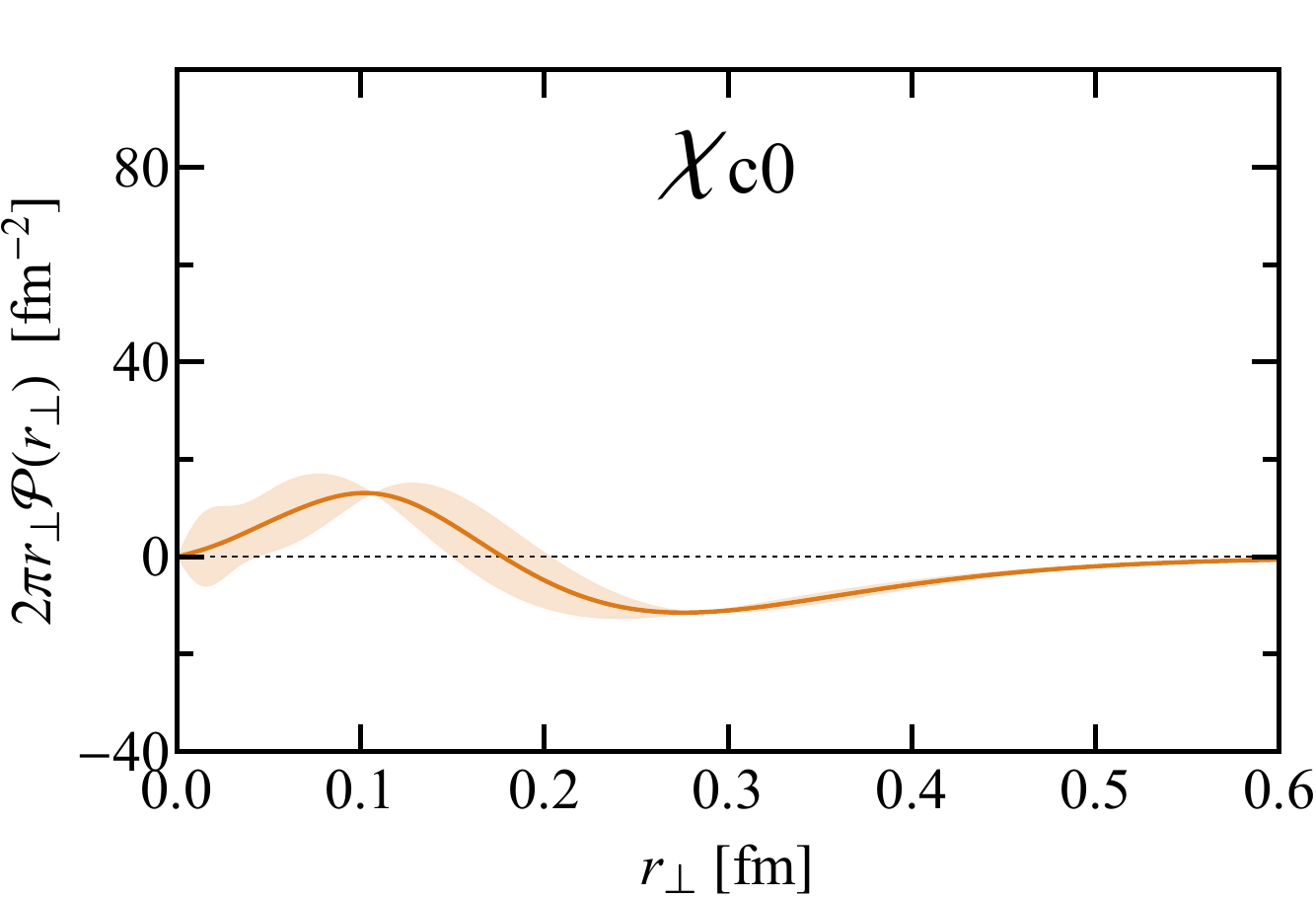}
\caption{Comparison of the pressure within $\eta_c$ (\textit{top}), $\eta_c'$ (\textit{middle}) and $\chi_{c0}$ (\textit{bottom}). The uncertainty bands are determined as described in the caption to Fig.~\ref{fig:etac_GFF_AD}.}
\label{fig:charmonia_pressure}
\end{figure}

From these form factors, we can extract physical densities. One of the advantages of the BLFQ approach is that the coordinate-space density can be obtained in closed form. This avoids the problem of numeric fitting or numerical Fourier transform, which introduces large and uncontrollable numerical errors. The extracted pressure is shown in Fig.~\ref{fig:charmonia_pressure} for $\eta_c$, $\eta'_c$ [i.e. $\eta_c(2S)$] and $\chi_{c0}$. The total pressure vanishes as a result of the von Laue condition. For all three states, the pressure is positive (repulsive) at the center of the meson and negative (attractive) near the edge, satisfying the stability conjecture \cite{Goeke:2007fp, Perevalova:2016dln}. The pressure profile of the radial excited charmonium $\eta'_c$ shows two additional nodes \cite{Mai:2012cx}. Note that the uncertainty of the pressure at small $r_\perp$ is large. This is because our basis has a finite UV coverage. The effective UV cutoff for $N_\text{max} = 8$ is $\Lambda_\text{UV} = \kappa \sqrt{N_\text{max}} \approx 2.8 \, \mathrm{GeV}$, corresponding to a spatial resolution of 0.07 fm. Smaller than this distance, the pressure starts to show large uncertainty, i.e. stronger basis dependence. 

\begin{figure}
\centering 
\includegraphics[width=0.475\textwidth]{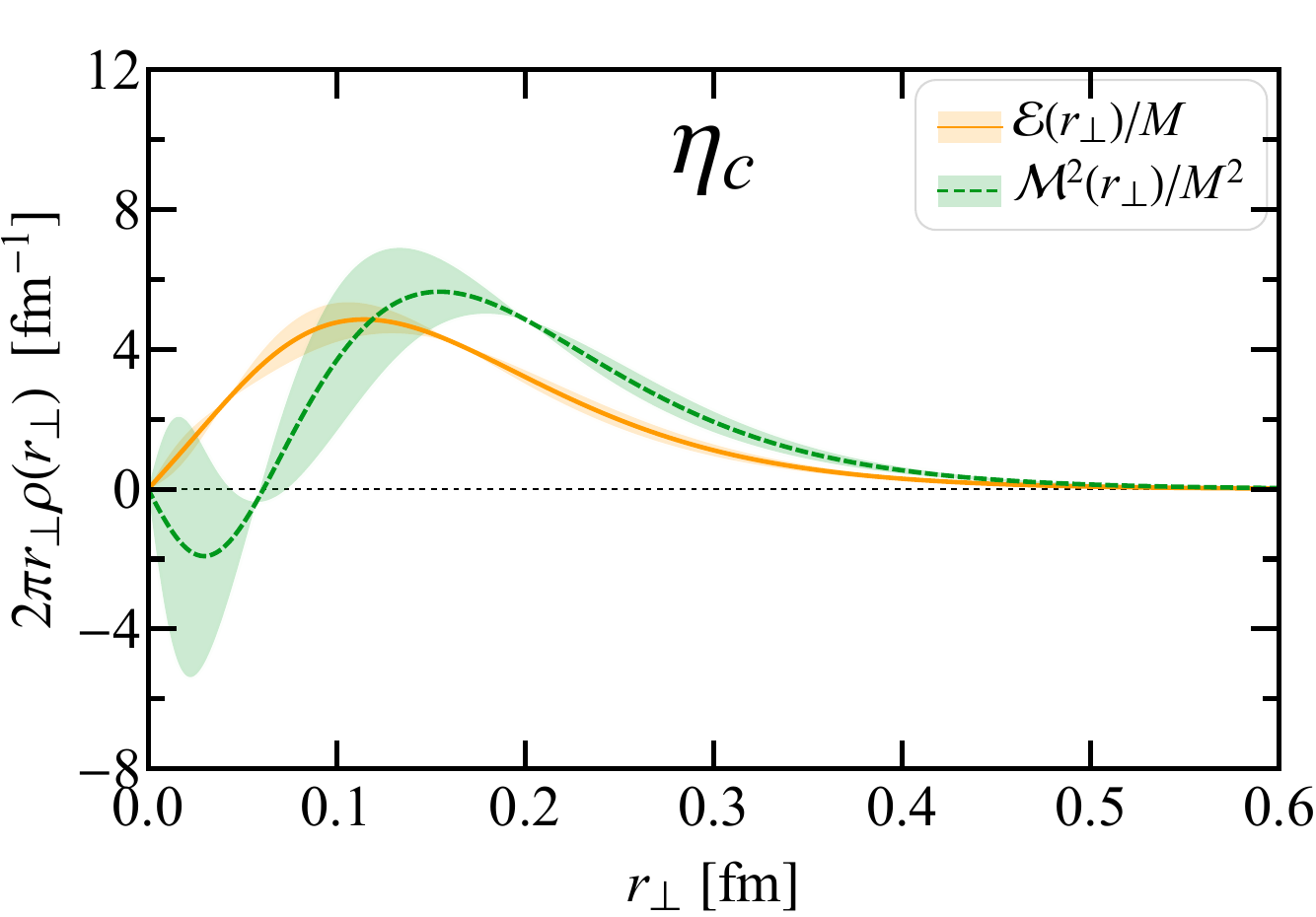}
\caption{Comparison of the normalized energy distribution and the normalized distribution of the invariant mass squared within $\eta_c$. The uncertainty bands are determined as described in the caption to Fig.~\ref{fig:etac_GFF_AD}. }
\label{fig:etac_energy}
\end{figure}

\begin{figure}
\centering 
\includegraphics[width=0.475\textwidth]{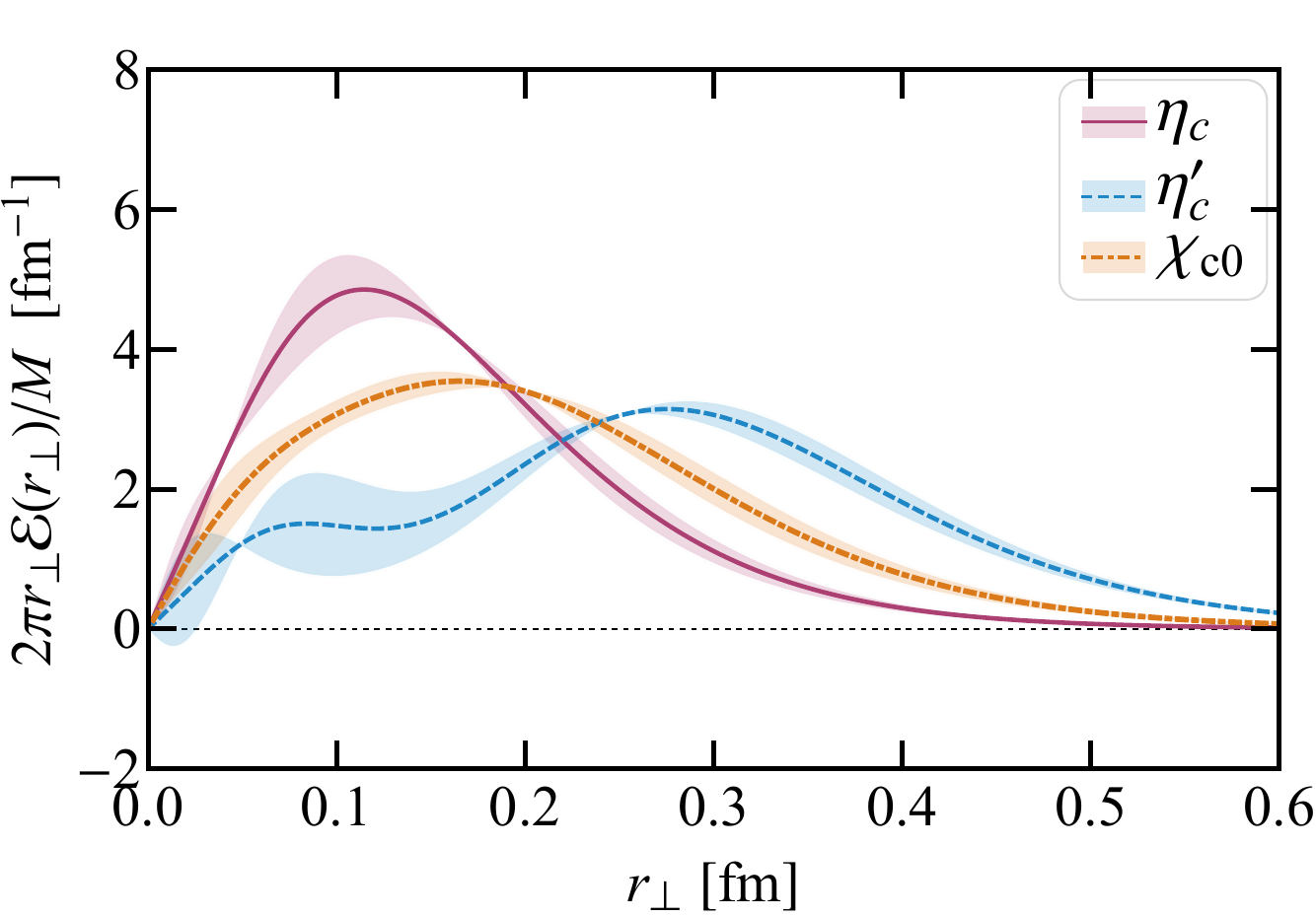}
\includegraphics[width=0.475\textwidth]{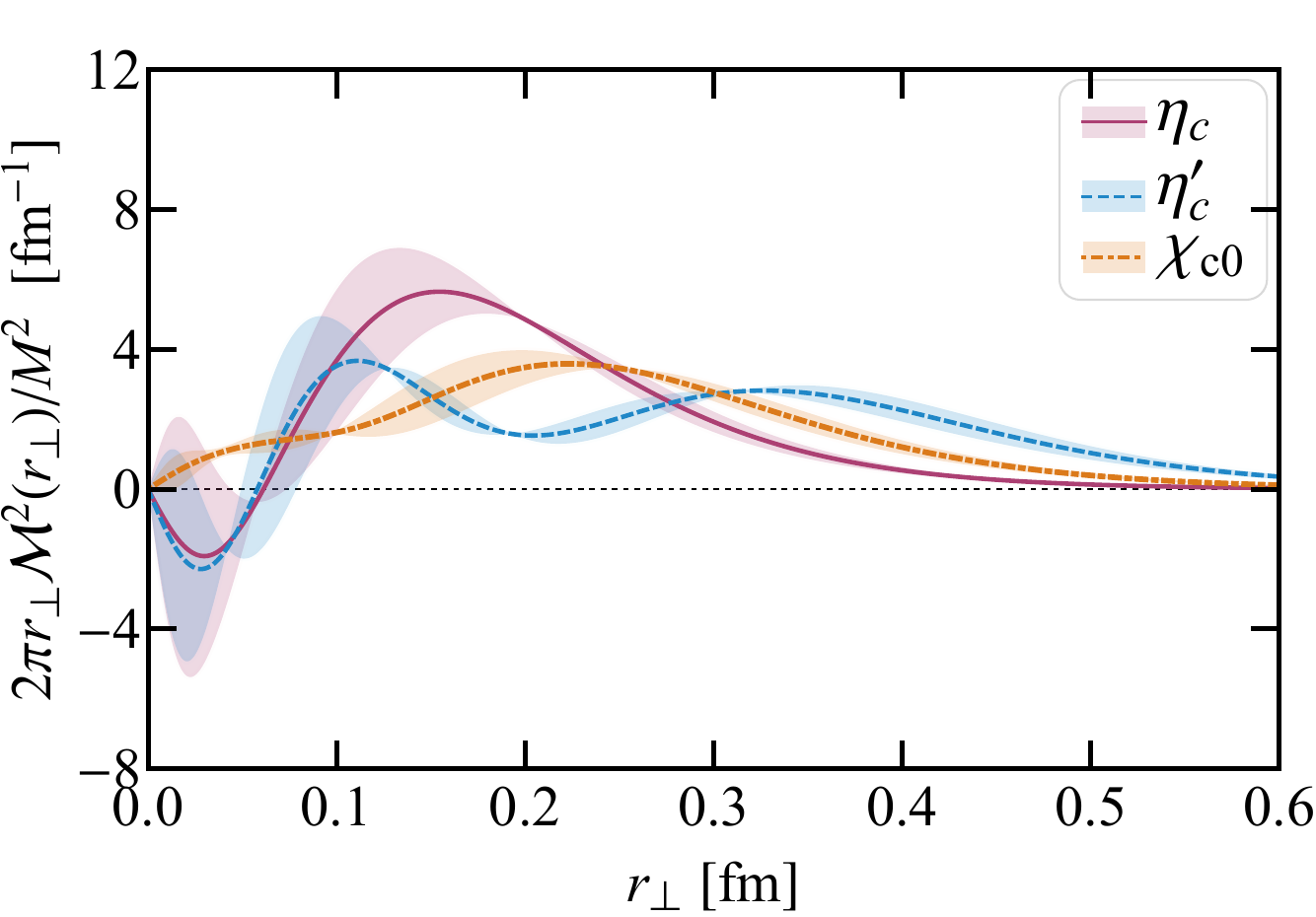}
\caption{Comparison of the energy distributions (\textit{top}) and the distribution of the invariant mass squared (\textit{bottom}) of $\eta_c'$ and $\chi_{c0}$. The uncertainty bands are determined as described in the caption to Fig.~\ref{fig:etac_GFF_AD}.}
\label{fig:charmonia_energy}
\end{figure}

We have mentioned the difference between the proper energy distribution $\mathcal E(r_\perp)$ defined from the instant form and the invariant mass squared distribution $\mathcal M^2(r_\perp)$ obtained on the light front. The former is normalized to the total energy of a hadron at rest, i.e. its rest mass $M$, whereas the latter is normalized to the invariant mass squared $M^2$ of the hadron. Fig.~\ref{fig:etac_energy} compares the normalized energy distribution of $\eta_c$ and its normalized invariant mass squared distribution. The energy distribution is positive, consistent with the weak energy condition. However, the invariant mass squared distribution becomes negative (i.e. tachyonic) at short distance $r_\perp$, although at the short distance, the results suffer from slow basis convergence as indicated by the wide bands.  Similar quantities can be extracted and compared for $\eta'_c$ and $\chi_{c0}$, as shown in Fig.~\ref{fig:charmonia_energy}. Within basis uncertainty, the energy distributions $\mathcal E(r_\perp)$ are consistent with the weak energy condition that the energy densities are positive. 

\begin{figure}
\centering 
\includegraphics[width=0.475\textwidth]{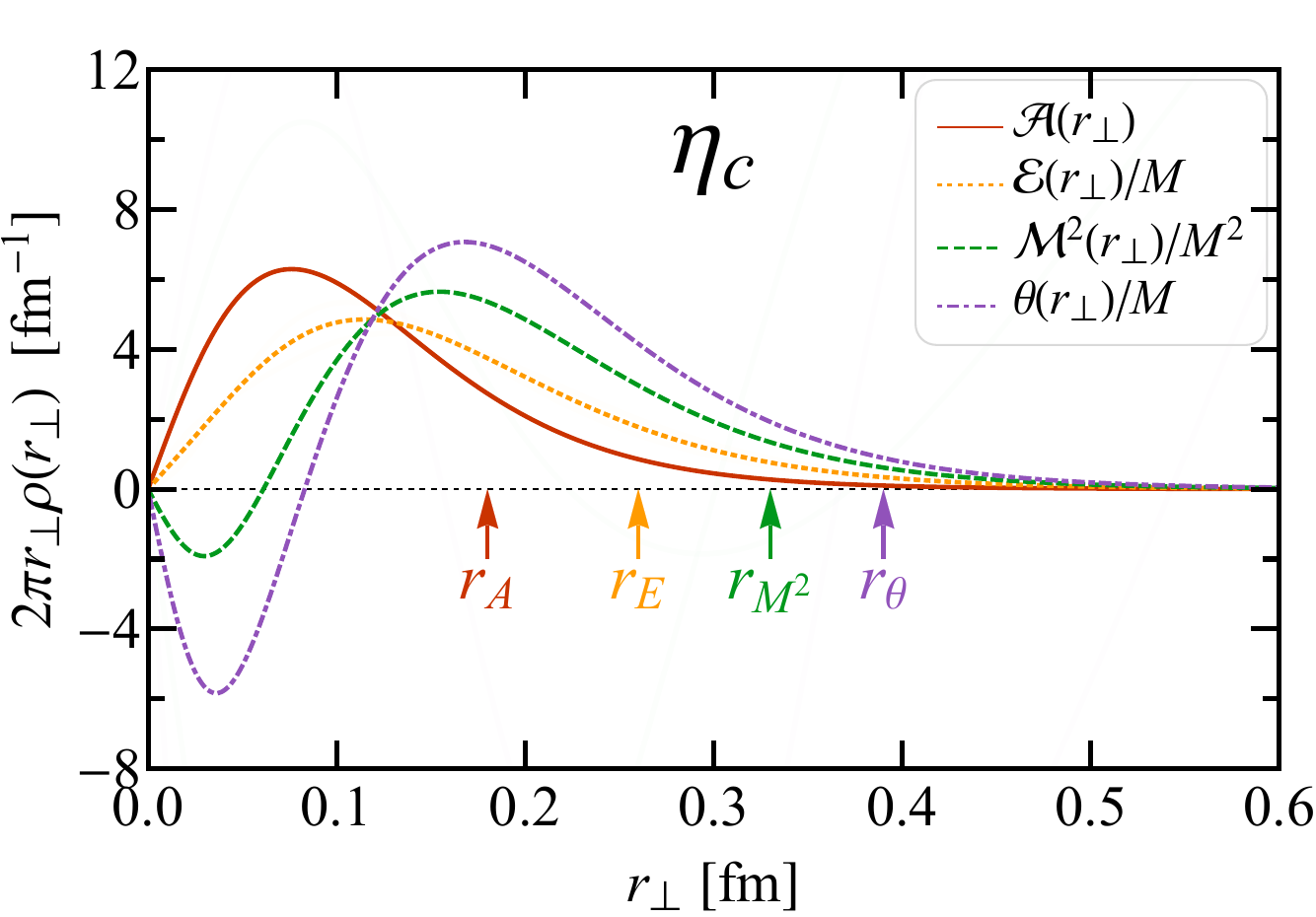}
\caption{Comparison of normalized densities of $\eta_c$. The densities are computed with LFWFs with $N_\text{max} = 8$ (see texts). The corresponding radii are indicated by arrows. }
\label{fig:etac_densities}
\end{figure}

We have so far introduced several densities: the matter density $\mathcal A(r_\perp)$, energy density $\mathcal E(r_\perp)$, invariant mass squared density $\mathcal M^2(r_\perp)$ and (trace) scalar density $\theta(r_\perp)$. Both $\mathcal E(r_\perp)$ and $\mathcal A(r_\perp)$ are strictly positive as required by the weak and null energy conditions. The energy conditions impose constraints on $D$ ({\ref{eqn:EC}}), which are satisfied by a negative $D$. 
We have argued that a negative $D$ suggests a hierarchy of the radii associated with these densities: 
\begin{equation}
r_A < r_E < r_{M^2} < r_\theta\,.
\end{equation}
Figure~\ref{fig:etac_densities} compares these densities for $\eta_c$ after the normalization, i.e. all densities are normalized to unity. Both the invariant mass squared density $\mathcal M^2(r_\perp)$ and the scalar density $\theta(r_\perp)$ become negative at the center of the meson. 

\begin{figure}
\centering 
\includegraphics[width=0.45\textwidth]{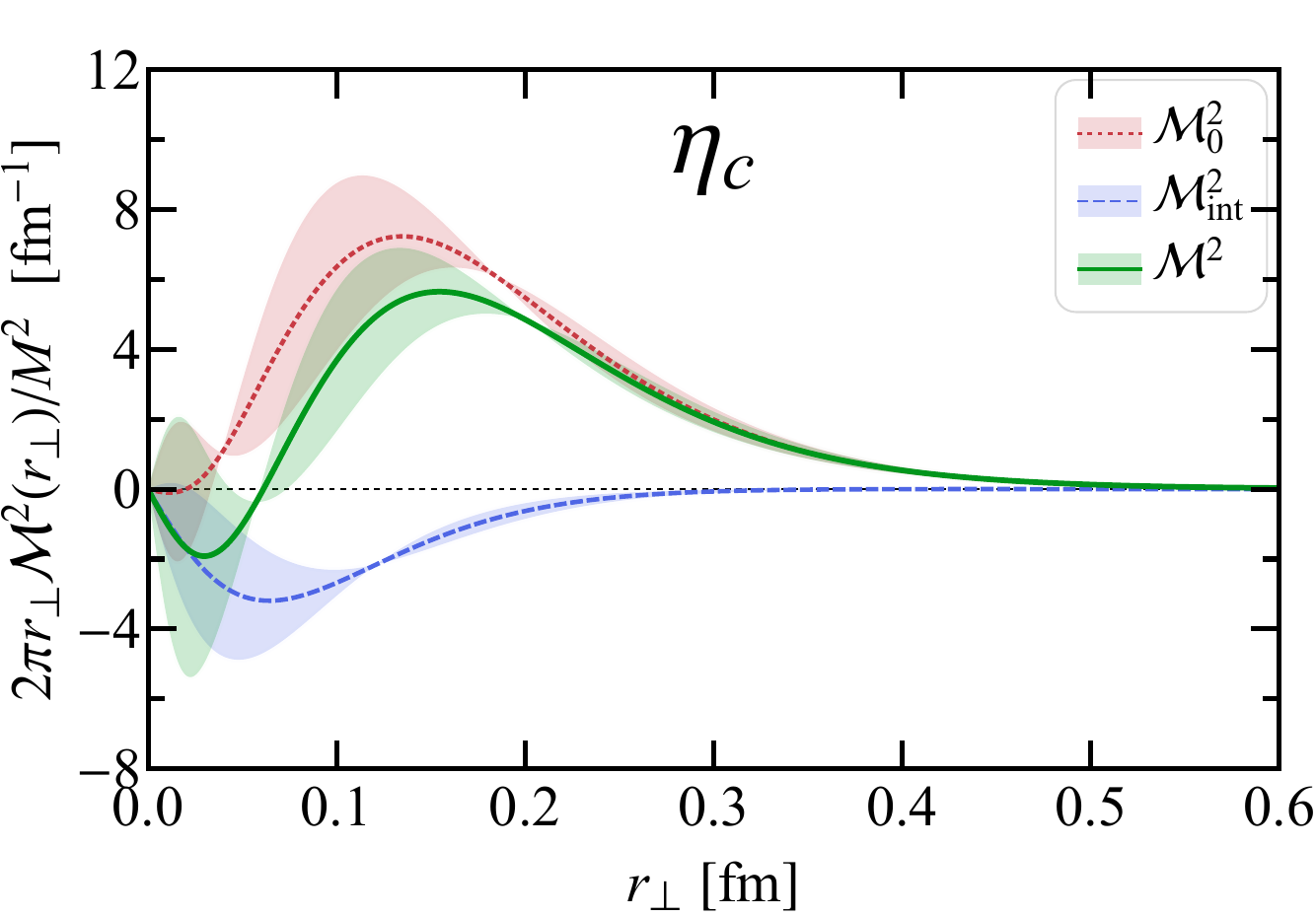}
\includegraphics[width=0.45\textwidth]{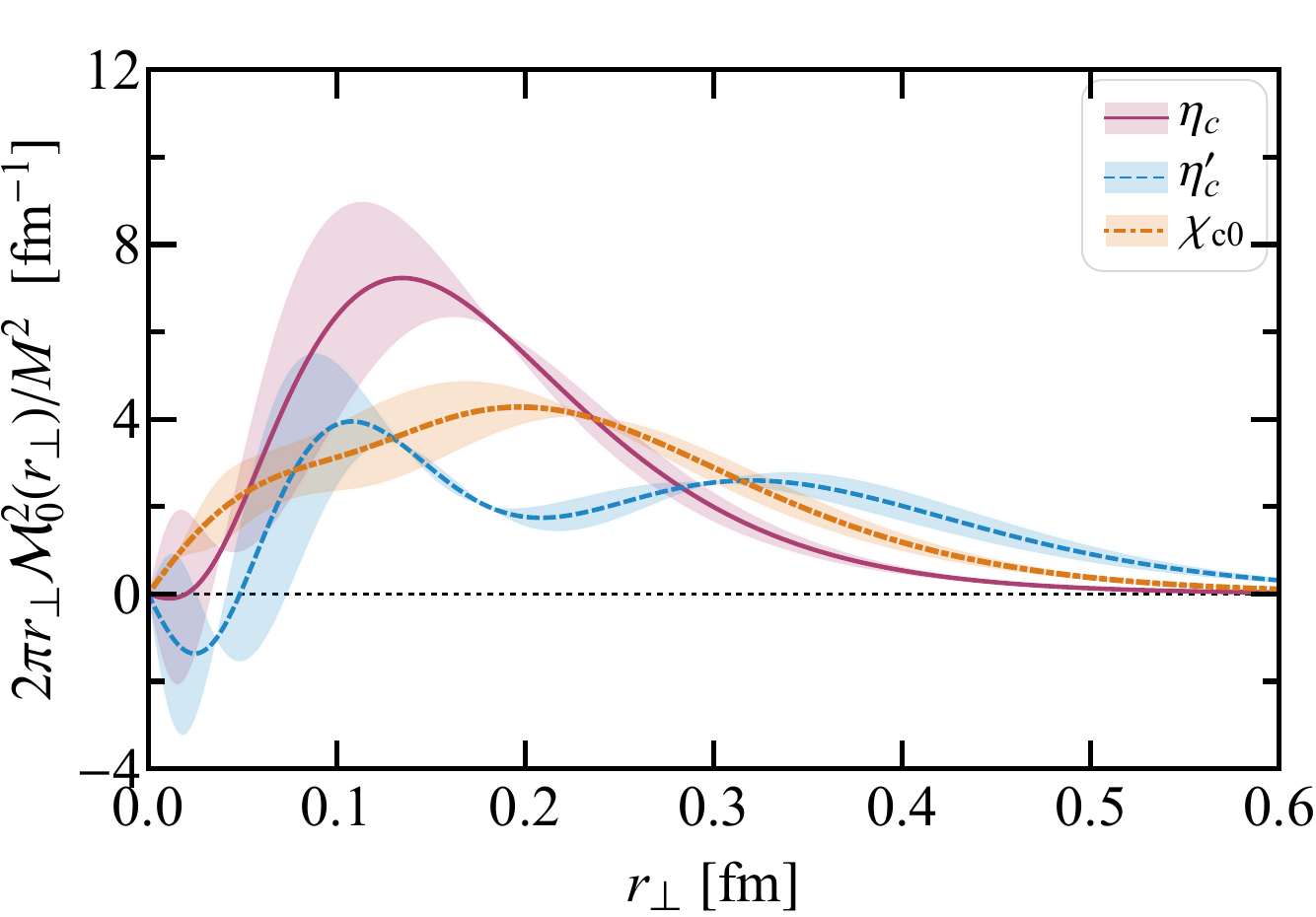}
\includegraphics[width=0.45\textwidth]{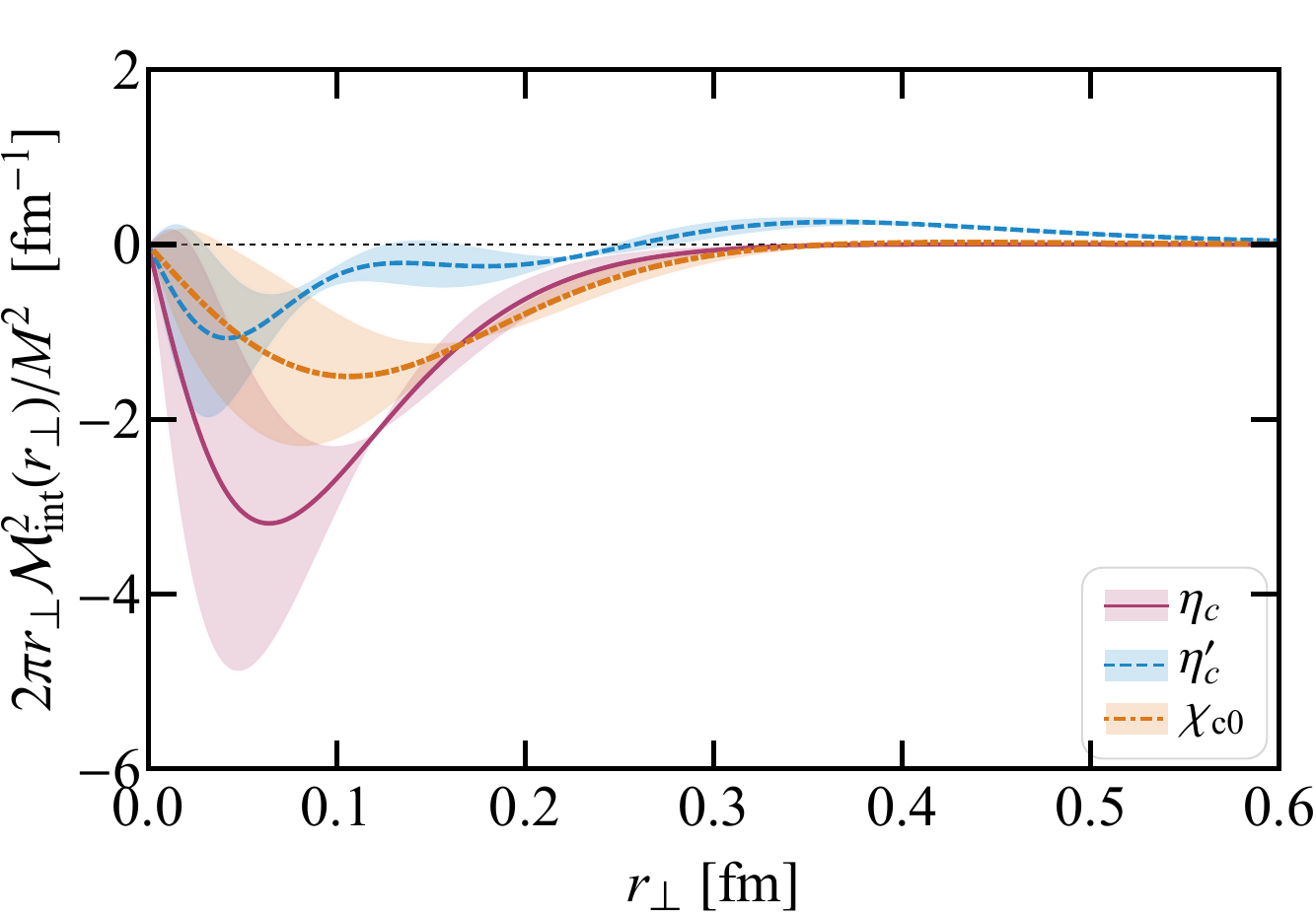}

\caption{(\textit{Top}) Decomposition of the internal light-front energy density, i.e. invariant mass squared density, of $\eta_c$ into the free part and the interacting part. (\textit{Middle}) Comparison of the free invariant mass squared densities for $\eta_c$, $\eta_c'$ and $\chi_{c0}$. 
(\textit{Bottom}) Comparison of the interacting invariant mass squared densities for $\eta_c$, $\eta_c'$ and $\chi_{c0}$.  The uncertainty bands are determined as described in the caption to Fig.~\ref{fig:etac_GFF_AD}.}
\label{fig:etac_decomposition}
\end{figure}

As we mentioned, in the light-front Hamiltonian formalism, we are also able to access the light-front kinetic energy and light-front potential energy separately. The distributions of the free internal light-front energy (i.e. free invariant mass squared) as well as the interacting light-front energy for $\eta_c$, $\eta_c'$ and $\chi_{c0}$ are shown in Fig.~\ref{fig:etac_decomposition}.  Note that $\mathcal M_0^2(r_\perp)$ does not remain positive for all distance for $\eta_c'$ and $\chi_{c0}$. This is caused by the recoil term $-{q^2_\perp}/{4x}$ in (\ref{eqn:M20}). The interaction density $\mathcal M_\text{int}^2(r_\perp)$ of $\eta_c'$ becomes positive at some large distance, consistent with the positive confining potential $r^2_\perp$ dominant at large parton separation. 

\begin{figure}
\centering
\includegraphics[width=0.35\textwidth]{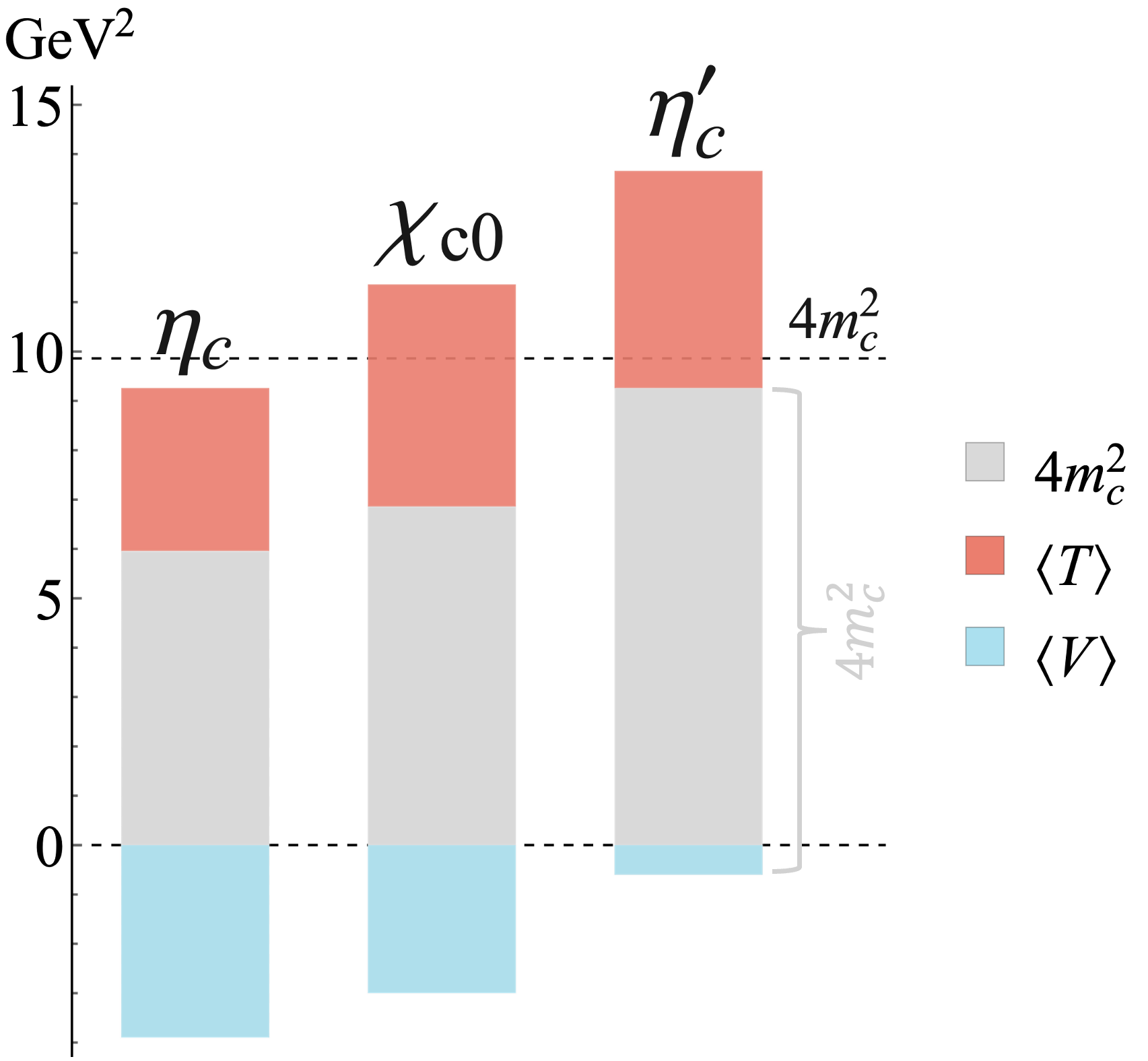}
\caption{Decomposition of charmonium invariant mass squared $M^2$ in terms of the total quark mass squared $4m_c^2$, kinetic light-front energy $\langle T \rangle$ and potential light-front energy $\langle V \rangle$. See texts. }
\label{fig:mass_decomposition}
\end{figure}

Non-relativistically, for particles within a potential of the form $V(r) = \alpha r^n$, the virial theorem\footnote{Lorcé et. al. generalized the virial theorem to QFT and showed that it is equivalent to von Laue condition \cite{Lorce:2021xku}.  } suggests, $n = {2\langle T \rangle}/{\langle V \rangle}$. Inspired by this result, we introduce a virial scaling index 
\begin{equation}
n \equiv \frac{2\langle T \rangle}{\langle V \rangle},
\end{equation}
where, we define the kinetic energy, i.e. the virial, as the free energy $M^2_0 \equiv M^2_0(Q^2=0)$ subtracting the square of the total quark mass,
\begin{equation}
\langle T \rangle \equiv M^2_0(Q^2=0) - (m_q + m_{\bar q})^2\,.
\end{equation}
The potential energy is simply the interacting energy,
\begin{equation}
\langle V \rangle \equiv M^2_\text{int}(Q^2=0)\,.
\end{equation}
This index $n$ characterizes the shape of the inter-particle force under spatial dilation. In non-relativistic limit, our effective interaction is a harmonic oscillator $r^2$ at large distance plus a Coulomb $r^{-1}$ at short distance. One would expect a virial index close to $-1$ for deep bound state, e.g. $\eta_c$, and $+1$ for bound state with large size \cite{Trawinski:2014msa}.  The obtained indices are collected Table~\ref{tab:D_radii}. Within the sizable basis uncertainty, these numbers are in agreement with the prediction from the non-relativistic virial theorem. 
Fig.~\ref{fig:mass_decomposition} shows the decomposition of the invariant mass squared $M^2$ of charmonia $\eta_c$, $\eta'_c$ and $\chi_{c0}$ into quark mass contribution, kinetic energy contribution and the potential energy contribution. 

\section{Summary and outlook}\label{sect:summary}

In this work, we investigate the gravitational form factors of the charmonium system. Starting from the light-front wave functions obtained previously in basis light-front quantization, we construct the wave function representation of GFFs $A$ and $D$. The former is extracted from the operator $T^{++}$ while the latter is from $T^{+-}$ with an impulse ansatz.
The obtained GFFs satisfy the known global constraints such as $A(0) = 1$, and the von Laue condition. 

From these two primary quantities, we also define physical densities on the light front, including the pressure $\mathcal P(r_\perp)$, the (proper) energy density $\mathcal E(r_\perp)$, invariant mass squared density (aka. internal light-front energy density) $\mathcal M^2(r_\perp)$ and  trace scalar density $\theta(r_\perp)$. We also identify the positive-defined density $\mathcal A(r_\perp)$, the Fourier transform of GFF $A(Q^2)$, as the (convective) matter density. The energy density $\mathcal E(r_\perp)$ is also positive and is consistent with the weak energy condition. 
These densities provide rich information of the system. For example, we find that there is a hierarchy for the sizes of the system: $r_A < r_E < r_{M^2} < r_\theta$, implying an onion-like structure of hadrons. 
Finally, we investigate the virial scaling index 
$2\langle T \rangle/\langle V \rangle$, which provides a quantitative measure of the strong force within charmonium.  The obtained values are consistent with the non-relativistic picture. 

We note that the method we proposed in this work is general enough and is applicable to states with a different spin such as $J/\psi$, and to other systems, such as the nucleon. The same method was applied to the pion in the context of holographic light-front QCD \cite{Li:2023izn}. It is interesting to observe the cancellation between the scalar and tensor glueballs and the emergence of scalar meson dominance within the pion. It is interesting to note that the scalar meson coupling to $D$ comes from the recoil term $-q^2_\perp/4x$. A closely related set of observables are the gravitational transition form factors \cite{Ng:1993vh, Ozdem:2019pkg, Polyakov:2020rzq, Alharazin:2020yjv, Kim:2022bwn, Ozdem:2022zig, Kim:2023xvw,  Alharazin:2023zzc, Alharazin:2023uhr}, which describe hadron production in gravity. These observables can also be accessed using the methods proposed in this work. 

One of the interesting questions is the comparison between GFF $D$ extracted from $T^{+-}$ and from $T^{12}$ and from $T^{11}+T^{22}$. Operators $T^{11}, T^{22}$ also contain interactions and need to be renormalized. However, for practical calculations on the light front, including perturbative or non-perturbative, counterterms computed from the light-front Hamiltonian are not sufficient to renormalize these operators. This is in sharp contrast to $T^{+-}$, which shares the same counterterms as its conserved charge, the light-front Hamiltonian. $T^{12}$ is interaction free. It is tempting to extract GFF $D$ from $T^{12}$ \cite{Nair:2024fit}. However, covariant light-front dynamics analysis shows that $T^{11}, T^{22}, T^{12}$ are all associated with a spurious GFF -- a structure that breaks the Poincaré symmetry. The GFF $D$ extracted from $T^{12}$ is likely to differ from what we obtained from $T^{+-}$, which does not suffer from the contamination of the spurious GFFs. 

Our model does not contain dynamical gluons or sea quarks. Therefore, we are not able to perform decomposition in terms of quarks and gluons. Incorporating dynamical gluons and sea quarks are necessary next steps in basis light-front quantization \cite{Lan:2021wok, Xu:2022abw, Xu:2023nqv, Lin:2023ezw, Kaur:2024iwn}. How to tame the computational complexity as well as the non-perturbative renormalization arising in these problems are critical challenges to be tackled. 

\section*{Acknowledgements}

The authors acknowledge fruitful discussions with C. Mondal, S. Nair, K. Serafin, J. More and C. Lorcé. 

 This work was supported in part by the National Natural Science Foundation of China (NSFC) under Grant No.~12375081, by the Chinese Academy of Sciences under Grant No.~YSBR-101, and by the US Department of Energy (DOE) under Grant No. DE-SC0023692. 
Y.L. is supported by the New faculty start-up fund of the University of Science and Technology of China. 
X. Z. is supported by new faculty startup funding by the Institute of Modern Physics, Chinese Academy of Sciences, by Key Research Program of Frontier Sciences, Chinese Academy of Sciences, Grant No.~ZDBS-LY-7020, by the Natural Science Foundation of Gansu Province, China, Grant No.~20JR10RA067, by the Foundation for Key Talents of Gansu Province, by the Central Funds Guiding the Local Science and Technology Development of Gansu Province, Grant No.~22ZY1QA006, by Gansu International Collaboration and Talents Recruitment Base of Particle Physics (2023-2027), by International Partnership Program of the Chinese Academy of Sciences, Grant No.~016GJHZ2022103FN, by National Natural Science Foundation of China, Grant No.~12375143, by National Key R\&D Program of China, Grant No.~2023YFA1606903 and by the Strategic Priority Research Program of the Chinese Academy of Sciences, Grant No.~XDB34000000.


\end{document}